\documentclass[preprint2,subfigure,epsfig]{aastex}
\onecolumn
 
\newcommand{\be}{\begin{eqnarray}}
\newcommand{\ee}{\end{eqnarray}}

\slugcomment{Submitted to The Astrophysical Journal}
\shortauthors{Demia\'nski \& Doroshkevich}
\shorttitle{The initial power spectrum}

\begin{document}
\title{Observational estimates of the initial power spectrum 
at small scale from Lyman-$\alpha$ absorbers}

\author{M. Demia\'nski\altaffilmark{1, 2, 3} and 
A. Doroshkevich\altaffilmark{3, 4} }
\altaffiltext{1}{Institute of Theoretical Physics, University 
of Warsaw, 00-681 Warsaw, Poland.} 
\altaffiltext{2}{Department of Astronomy, Williams 
College, Williamstown, MA 01267, USA}
\altaffiltext{3}{Theoretical Astrophysics Center, Juliane Maries 
Vej30, DK-2100,  Copenhagen, Denmark.}
\altaffiltext{4}{Keldysh Institute of Applied Math. Russian 
Academy of Sciences, 125047, Moscow,  Russia}

\begin{abstract}
We present a new method of measuring the power spectrum of 
initial perturbations to an unprecedently small scale of 
$\sim$ 10$h^{-1}$ kpc. We apply this method to a sample 
of 4500 Ly-$\alpha$ absorbers and recover the cold dark matter 
(CDM) like power spectrum at scales $\geq 300h^{-1}$kpc with 
a precision of $\sim$ 10\%. However at scales $\sim 10 - 300 
h^{-1}$kpc the measured and CDM--like spectra are  noticeably 
different. This result suggests a complex inflation with 
generation of excess power at small scales. The magnitude 
and reliability of these deviations depend upon the 
possible incompleteness of our sample and poorly understood 
process of formation of weak absorbers. 
Confirmation of the 
CDM--like shape of the initial power spectrum or detection 
of its distortions at small scales are equally important 
for widely discussed  problems of physics of the early 
Universe, galaxy formation, and reheating  of the Universe. 

We use the Zel'dovich theory of gravitational instability 
to derive  statistical description of the properties of observed 
structure. Our method links the observed mass function of 
absorbers with the correlation function of the initial velocity 
field and therefore it avoids the Nyquist restrictions limiting 
the investigations based on the smoothed flux or density fields. 
This approach is in general consistent with numerical simulations 
of the process of structure formation, describes reasonably well 
the Large Scale Structure observed in the galaxy distribution 
at small redshifts and emphasizes the generic similarity of 
galaxies and absorbers.

The physical model of absorbers adopted here asserts that 
they are formed in the course of both linear and nonlinear 
adiabatic or shock compression of dark matter (DM) and 
gaseous matter. It allows us to link the column density 
and overdensity of DM and  gaseous components with observed 
characteristics of  absorbers such as the column density 
of neutral hydrogen,  redshifts and Doppler parameter. At 
scales $\geq 1h^{-1}$Mpc all characteristics of the DM 
component and, in particular,  their redshift distribution 
are found to be consistent with theoretical expectations 
for Gaussian initial perturbations with a CDM--like power 
spectrum. 
\end{abstract} 
\maketitle

\twocolumn

\section{Introduction}

One of the most important problems of modern cosmology 
is to determine the power spectrum of primordial perturbations. 
This problem is closely connected with physics of the early 
Universe, properties of the inflation process, reionization 
of the Universe at moderate redshifts and formation of the 
Large Scale Structure (LSS) observed in deep galaxy surveys 
among others. 

Presently various observational approaches are used to measure 
the power spectrum of the initial density perturbations. The 
amplitude and the shape of initial power spectrum on scales 
$\geq 10h^{-1}$Mpc are approximately established by 
investigations of the relic radiation (Spergel et al. 2003) 
and the structure  of the Universe at $z\ll$~1 detected in 
large redshift surveys (Percival et al. 2001; Peacock et al. 
2001; Efstathiou et al. 2001; Tegmark, Hamilton \& Xu 2002;  
Verde et al. 2002) and weak lensing data (see, e.g., Hoekstra, 
Yee \& Gladders 2002). The shape of the initial power spectrum 
on scales $10h^{-1}$Mpc -- $1h^{-1}$Mpc can be tested at high 
redshifts where it is not yet strongly distorted by nonlinear 
evolution (Croft et al. 1998, 2002; Nusser \,\&\,Haehnelt 2000; 
Gnedin \& Hamilton 2002). Recent results on reconstruction of
the initial power spectrum are summarized and discussed in 
Tegmark and Zaldarriaga (2002), Wang et al. (2002) and Peiris 
et al. (2003).  

Analysis of the observed absorption spectra of high redshift 
quasars seems to be the most promising method of recovering 
the power spectrum of initial perturbations at small scale.  
Indeed, the absorption lines trace the small scale hydrogen
distribution along the line of sight at redshifts $z\geq$ 2 
when matter is not yet strongly clustered. The available 
Keck and VLT high resolution observations of the Lyman$-\alpha$ 
forest provide a sufficiently rich database which can be 
analyzed with the help of statistical methods. 

The composition and spatial distribution of the observed 
absorbers is complicated and at low redshifts a significant 
number of stronger Ly-$\alpha$ lines and metal systems is 
associated with galaxies (Bergeron et al. 1992; Lanzetta et 
al. 1995; Tytler 1995; Le Brune et al. 1996). However as was 
recently shown by Penton, Stock and Shull (2002), even at 
small redshifts some absorbers are associated with galaxy 
filaments while others are found within galaxy voids. 

These results suggest that the population of weaker absorbers 
dominating at higher redshifts can be associated both with 
the periphery of rare high density objects and with weaker DM 
pancakes formed by the non luminous baryonic and DM components 
in extended low density regions. They suggest that the 
Ly--$\alpha$ forest can be considered as a low mass component 
of the generic LSS which is seen in simulated and observed 
spatial matter distribution. In turn, the relatively homogeneous 
spatial distribution of absorbers implies a more homogeneous 
spatial distribution of both DM and baryonic matter as compared 
with the observed distribution of the luminous matter.

Essential progress in understanding the process of formation and 
evolution of LSS has recently been achieved through numerous high 
resolution simulations which show that this process is mainly 
driven by the power spectrum of initial perturbations. These 
results were used in Demia\'nski \& Doroshkevich (1999, 2002; 
hereafter DD99 \& DD02) to consider the properties of the LSS 
in the context of the nonlinear theory of gravitational 
instability (Zel'dovich 1970; Shandarin \& Zel'dovich 1988). The  
statistical description of the process of structure formation 
and evolution developed in DD99 \& DD02 links the directly 
measured characteristics of the structure such as the mass 
function of DM pancakes and walls, with the initial power 
spectum. 

This approach generalizes the Press--Schechter formalism (Press \& 
Schechter 1974; Peacock \& Heavens 1990; Bond et al. 1991; Loeb \& 
Barkana 2001) and describes the formation and evolution of all DM 
structure elements for the CDM or WDM initial power spectra without 
any smoothing or filtering procedures. The evolution of structure 
proceeds through random formation and merging of Zel'dovich 
pancakes, their transverse expansion and/or compression and 
subsequent transformation into high density clouds and filaments. 
Later on the hierarchical merging of pancakes, filaments and 
clouds forms rich galaxy walls observed at small redshifts. The 
main stages of this evolution are driven by the initial power
spectrum. 

As is well known, the Zel'dovich approach correctly describes both 
the linear and mildly nonlinear stages of the structure formation 
but it cannot describe the processes of violent relaxation and 
the final stage of formation of gravitationally bound object. 
In spite of this, the statistical approach proposed in DD99 
and DD02 nicely describes the main properties of observed and 
simulated LSS (Demia\'nski et al. 2000; Doroshkevich, Tucker \& 
Allam 2002). In particular, comparison of characteristics 
of simulated DM distribution and observed galaxy walls with 
theoretical expectations allows one to restore the amplitude 
and shape of initial power spectrum at appropriate scales.

Early application of the {\it truncated} Zel'dovich 
approximation for the description of the Ly-$\alpha$ forest 
was critically mentioned in Bond and Wadsley (1997) but their 
criticism was not supported by more detailed investigation. 
In contrast, the Zel'dovich approach has been considered
as very promissing by Hui, Gnedin and Zhang 
(1997) and Hui (1997) who applied it to describe  
the Doppler parameter and neutral hydrogen column density. 
However, these characteristics depend upon random variations 
of the gas entropy and, so, their theoretical description is 
quite problematic. 

The statistical approach proposed in DD02 was used in 
Demia\'nski, Doroshkevich \& Turchaninov (2003, hereafter 
Paper I) to describe properties of absorbers observed in 
14 high resolution spectra. This approach deals with DM 
charactristics of absorbers. Evolution of DM component 
of absorbers is driven 
mainly by the gravitational interaction and it can be 
approximately described theoretically. However in this 
approach the DM characteristics of absorbers are expressed 
through the observed ones what in turn requires a physical 
model of absorbers.

Such model, introduced in Paper I, allows one to find a physically
motivated combination of observed characteristics of absorbers which
are consistent with theoretically expected ones for the DM
pancakes. These results allows us to interpret this combination as
actual characteristic of DM component of absorbers and confirm the
self consistency of the proposed physical model of absorbers.  They
show also that the statistical approach proposed in DD99 and DD02
provides a reasonable description at least for so defined DM
characteristics of absorbers. However, we cannot theoretically
describe more complex evolution of the gaseous component of absorbers,
their Doppler parameters and column density of neutral hydrogen.

Following Paper I, also in this paper we consider that absorbers are  
dominated by long--lived gravitationally bound and partially relaxed 
objects formed in the course of both adiabatic and shock compression. 
Observations of galaxies and quasars at $z\geq 3$ and the 
reheating of the Universe at redshifts $z\geq 6$ demonstrate 
the importance of nonlinear processes at such redshifts and 
support our approach. Numerous simulations (see, e.g. Frenk 
2002) confirm the formation of high density DM filaments and 
sheets at similar redshifts ($z\geq 2$). The typical size and 
mass of such objects progressively increase with time and now 
richer objects are seen as the LSS in the  galaxy distribution. 
In this paper we assume that absorbers at high $z$ trace the 
DM structure which is qualitatively similar to the rescaled 
one observed at small redshifts. However, for less massive 
pancakes properties of the baryonic and DM components differ 
due to the influence of gaseous pressure. 

Here we use this approach to recover the initial power spectrum 
to unprecedently small scales. In contrast with previous 
investigations (Croft et al. 1998, 2002; Nusser\,\&\,Haehnelt 
2000) we analyse the DM mass function of absorbers rather than 
the flux or smoothed density field. This means that our results 
are not restricted by the standard factors such as the Nyquist 
limit, the impact of nonlinear processes, the unknown matter 
distribution between absorbers or their peculiar velocities. 
The reliability and precision of our method have been tested  
on the simulated DM and observed galaxy surveys 
(Demia\'nski et al. 2000; Doroshkevich, Tucker \& Allam 2002). 

At scales $\geq 1 h^{-1}$Mpc our estimates of the initial power 
spectrum are consistent with the CDM--like one with precision 
$\sim$ 10\%. However, we found some evidence that at scales 
$\leq 1 h^{-1}$Mpc the initial power spectrum differs from the 
CDM--like one suggesting a complex inflation with generation of 
excess power at small scales. Such excess power accelerates the 
process of galaxy formation at high redshifts and shifts the 
epoch of reheating of the Universe, which perhaps began at 
$z=20\pm 10$ (Kogut et al. 2003) and continued up to 
$z\approx$~6 (Djorgovski et al 2001; Fan et al. 2001, 2003). 
More detailed discussion of the process of reionization can be 
found in Ciardi, Ferrara \& White (2003) and Cen (2003).

At present we have only limited information about the properties 
of the background gas and UV radiation (Scott et al. 2000; Schaye 
et al. 2000; McDonald \& Miralda--Escude 2001; McDonald et al. 
2000, 2001; Theuns  et al. 2002 a, b) and therefor some numerical 
factors in our model remain undetermined. This means that our 
approach should be tested on representative  numerical simulations 
which more accurately follow the process of formation and disruption 
of pancakes and filaments and provide a unified picture of the 
process of absorbers formation and evolution (see, e.g., Weinberg
et al. 1998; Zhang et al. 1998; Dav\'e et al. 1999; Theuns et al.
1999). However, as was shown by Meiksin, Bryan \& Machacek (2001), 
available simulations reproduce quite well the characteristics of the 
flux but cannot restore other observed characteristics of the 
forest. This means that first of all numerical simulations 
should be improved (see Sec. 9.4). 

This paper is organized as follows. In Secs. 2 \& 3 a short
theoretical description of structure formation and the physical 
model of absorbers are presented. In Sec. 4 the observational 
databases used in our analysis and redshift evolution of the 
observed absorber characteristics are described.  Results of 
statistical analysis are given in Secs. 5 and 6 and their 
reliability and precision are discussed in Sec. 7. In Sec. 8 
we link the measured correlation function with the power 
spectrum. Discussion and conclusions can be found in Sec. 9.

\section{Initial power spectrum and characteristics of DM 
pancakes}

\subsection{Basic cosmological model}

In this paper we consider the spatially flat $\Lambda$CDM 
model of the Universe with the Hubble parameter and mean 
matter density given by:
\be
H^{2}(z) = H_0^2\Omega_m(1+z)^3[1+\Omega_\Lambda/\Omega_m
(1+z)^{-3}]\,,
\label{basic}
\ee
\[
\langle\rho_m(z)\rangle = {3H_0^2\over 8\pi G}\Omega_m(1+z)^3,
\quad H_0=100h\,{\rm km/sMpc}\,.
\]
Here $\Omega_m=0.3\,\&\,\Omega_\Lambda=0.7$ are dimensionless 
matter density and the cosmological constant (dark energy), 
and $h=$~0.7 is the dimensionless Hubble constant. 

We assume that the baryonic background is characterized by the
following parameters 
\be
\langle n_b(z)\rangle = 2.4\cdot 10^{-7}{\rm cm}^{-3}
(1+z)^3(\Omega_bh^2/0.02)\,,
\label{bg}
\ee
\[
T_{bg}\approx 1.6\cdot 10^4K,\quad b_{bg}=
\sqrt{2k_BT_{bg}\over m_H}\approx 16{\rm km/s}\,.
\]
Here $\Omega_b$, $T_{bg}\,\&\,b_{bg}$ are the  
dimensionless density, temperature and Doppler parameter 
of the baryonic component, $k_B\,\&\,m_H$ are the Boltzmann 
constant and the mass of hydrogen atom.

\subsection{The initial power spectrum}

As a reference power spectrum of initial perturbations we take 
the standard CDM--like spectrum with the Harrison -- Zel'dovich 
asymptotic,  
\be
P(k)={A^2k\over 4\pi k_0^4}T^2(k/k_0)D_W(k)D_J(k),\quad 
k_0={\Omega_mh^2\over{\rm Mpc}}\,,
\label{f1}
\ee
where $A$ is the dimensionless amplitude of perturbations, 
$k$ is the comoving wave number. The transfer function, $T(k)$, 
and the damping factor $D_W(k)$ describing the free streaming 
of DM particles were given in Bardeen et al. (1986). The factor 
$D_J(k)$ describes the dumping of small scale baryonic 
perturbations (Jeans scale) and the bias between perturbations 
of DM and baryonic components (see, e.g., Mattarese \& Mohayaee 
2002). However, characteristics of the DM structure elements 
are defined by (\ref{f1}) with $D_J(k)\equiv 1$. 

For WDM particles the dimensionless damping scale, $R_f$, 
and the damping factor, $D_W$, are
\[
R_f={1\over 5}\left({\Omega_mh^2{\rm keV}}\over M_{DM}
\right)^{4/3},\quad D_W=\exp[-\eta R_f-(\eta R_f)^2]\,,
\]
where $\eta=k/k_0$ and $M_{DM}$ is the mass of WDM particles 
(Bardeen et al. 1986). The Jeans wave number, $k_J$, and the 
damping factor, $D_J$, can be taken as  
\[
k_J^{-1}\approx 0.7 b_{bg}(1+z)H^{-1}(z),\quad 
D_J\approx (1+\eta_J^2\eta^2)^{-1}\,,
\]
\[
\eta_J=k_0/k_J\approx 2\cdot 10^{-2}\sqrt{\Omega_mh^2\over
0.13}\sqrt{4\over 1+z}\,{b_{bg}\over 16{\rm km/s}}\,. 
\]
Here as a reference we use the values of the background 
parameters specified by (\ref{bg}).

The amplitude of initial perturbations, $A$, is now
measured reasonably well with different methods.  
It is simply linked with $\sigma_8$, the variance of the 
mass within a randomly placed sphere of radius 8$h^{-1}$Mpc,  
\be
\sigma_8^2 = 9A^2\int_0^\infty d\eta \eta^3T^2(\eta)
\left({\sin \eta_8-\eta_8\cos \eta_8\over \eta_8^3}
\right)^2\,,
\label{as8}
\ee
where $\eta=k/ k_0$ and $\eta_8=8\eta\Omega_mh$. Here 
we use the latest estimates (Spergel et al. 2003) for 
the model (\ref{basic})
\be
\sigma_8\approx 0.055 A\approx 0.9\pm 0.1,\quad A\approx 
16.4\pm 1.82\,.
\label{s8}
\ee

\subsection{Coherent lengths and correlation functions 
of the initial perturbations}

For the spectrum (\ref{f1}) the coherent lengths of velocity 
and density fields, $l_v\,\&\,l_\rho$, are expressed through 
the spectral moments, $m_{-2}\,\&\,m_0$, as follows: 
\be
l_v={1\over k_0\sqrt{m_{-2}}}={6.6\over \Omega_mh^2}
{\rm Mpc}\approx 33.8 h^{-1}{\rm Mpc}{0.13\over
\Omega_mh^2} \,, 
\label{lv}
\ee
\[
m_{-2} = \int_0^\infty d\eta~\eta T^2(\eta)D_W(\eta)D_J(\eta)
\approx 0.023,\quad l_\rho=q_0 l_v\,,
\]
\[
m_0 = \int_0^\infty d\eta~\eta^3 T^2(\eta)D_W(\eta)D_J(\eta),
\quad q_0=5{m_{-2}^2\over m_0}\,.
\]

The analysis of the redshift distribution of Lyman-$\alpha$ 
clouds (Paper I) allowed us to estimate the moment $m_0$ that 
in turn restricts the expected Jeans scale and the mass of 
the dominant fraction of DM particles. These estimates 
\be
q_0\approx(0.6 - 1.2)\cdot 10^{-2},\quad 
\eta_J\approx 0.01-0.02\,,
\label{qmm}
\ee
\[
M_{DM}\approx (1.5 - 5){\rm keV},\quad R_f\approx (7.7 
- 1.5)\cdot 10^{-3}\,,
\]
are close to those of Narayanan et al. (2000), $M_{DM}\geq$ 
0.75 keV, and Barkana, Haiman \& Ostriker (2001), $M_{DM}
\geq$~1 -- 1.25~keV. However, as was discussed in Paper I, the 
Jeans damping, $D_J$, restricts abilities of this approach, 
and to test the small scale power spectrum we have to use 
characteristics of DM component of absorbers for which 
$D_J=1$\,. 

As was demonstrated in DD99 and DD02, main statistical 
characteristics of structure are expressed through the 
normalized longitudinal correlation function of the initial 
velocity field  
\be
\xi_v(q)=3{\langle({\bf q\cdot v}(\tilde{\bf q}_1))
({\bf q\cdot v}(\tilde{\bf q}_2))\rangle\over 
\sigma_v^2q^2},\quad{\bf q}={\tilde{\bf q}_1-\tilde{\bf 
q}_2\over l_v}\,.
\label{xiv0}
\ee
Here $\tilde{\bf q}_1 ~\&~ \tilde{\bf q}_2$ are the  
unperturbed coordinates of two particles at $z=$ 0, 
$q=|{\bf q}|$, and $\sigma_v^2$ is the velocity variance. 
This function is expressed through the power spectrum: 
\be    
\xi_v={3\over m_{-2}}\int_0^\infty d\eta\,\eta^2\cos x
\int_\eta^\infty {dy\over y^2}T^2(y)D_W(y) = 
\label{xivv}
\ee
\[
{\sqrt{\pi/2}\over m_{-2}}\int_0^\infty {d\eta\over
\sqrt{x}}[J_{1/2}(x)-2J_{5/2}(x)]\eta T^2(\eta)D_W\,,
\]
where $x=q\eta,\, J_{1/2}(x)$ and $J_{5/2}(x)$ are Bessel 
functions. 
From (\ref{xivv}) it follows that  $\xi_v(0)=1$ and $\int_0^\infty 
dq~\xi(q)=0$. For the CDM -- like spectrum (\ref{f1}) and for the 
most interesting ranges $q_0\leq 10^{-2}$, $0.5\geq q\geq 10^{-3}$, 
the velocity correlation function can be fitted as follows:       
\be       
\xi_v=\xi_{CDM}\approx 1-{q^2\over\sqrt{u_0(q) u_1(q)}}{q_0+1.5
u_0(q)\over q_0+u_0(q)}\,.       
\label{xiv}       
\ee 
\[
u_0=\sqrt{q^2+q^2_0},\quad u_1=\sqrt{q^2+3\cdot 10^{-4}}
\]
Further on, we will use this function with $q_0=0.01$ as the 
reference one and will compare it with observational estimates 
of $\xi_v(q)$. This value of $q_0$ lies in the range of 
(\ref{qmm}) and corresponds to the mass of WDM particles 
$M_{WDM}\approx$ 3 keV (or the Jeans damping scale for
 $T_{bg}\approx 10^4$K).
     
\subsection{Characteristics of DM pancakes}       

We use the Zel'dovich approximation (Zel'dovich 1970; Shandarin 
\& Zel'dovich 1989) to relate the characteristics of the initial 
density perturbations with the properties of observed pancakes. 
As was discussed in DD99, DD02 and Paper I, three quantitative 
characteristics of pancakes can be used for the comparison with 
observations. Two of them are the differential, $N_q$, and 
cumulative, $W_q$, distribution functions (PDFs) of the 
dimensionless DM surface density (or column density) of 
pancakes, $q$. The DM surface density is defined as the 
dimensionless unperturbed distance at redshift $z=0$ between 
DM particles bounding the pancake (\ref{xiv0}). The third 
characteristic is the dimensionless mean number density of 
absorbers (Lyman-$\alpha$ clouds), $n_{abs}(q_{thr},z)$. All 
these characteristics depend upon the distribution function 
of the initial density perturbations which can be taken as 
Gaussian (Komatsu et al. 2003), upon the correlation function 
of the initial velocity field, $\xi_v(q)$, and the survival 
probability of pancakes, $W_s(q)$, (DD99, DD02). The survival 
probability of clouds was introduced by Peacock \& Heavens 
(1990) and Bond et al. (1991) to describe, in the framework 
of Press--Schechter approach, the difference between the 
fraction of matter accumulated by clouds and their mass 
function.

As was shown in DD99 and DD02, both PDFs and the survival 
probability depend upon the self similar variable
\be
\zeta = {q^2\over 4\tau^2(z)[1-\xi_v(q)]}\,,
\label{xit}
\ee 
where the 'time' $\tau(z)$ describes the growth of 
perturbations due to the gravitational instability. 
For the $\Lambda$CDM cosmological model (\ref{basic}), 
and for $z\geq$ 2 we have 
\be
\tau(z)\approx \tau_0\left({1+1.2\Omega_m\over 2.2\Omega_m}
\right)^{1/3}{1\over 1+z}\approx {1.27\tau_0\over 1+z}\,,
\label{Btau}
\ee
\be
\tau_0=Am_{-2}/\sqrt{3}\approx 0.22\pm 0.02\,, 
\label{tau0} 
\ee 
where $A$ was introduced in (\ref{f1}) and (\ref{s8}).

For the Gaussian initial perturbations the matter fraction 
accumulated by structure elements is proportional to 
erfc$(\zeta)$=1-erf$(\zeta)$, where erf$(\zeta)$ is the 
standard error function. Following Peacock \& Heavens (1990) 
and  Bond  et al. (1991), we assume 
that the mass function of the  DM column density of pancakes 
and the matter fraction accumulated  by pancakes differ by 
the survival probability, $W_s(\zeta)$. For low mass pancakes 
$W_s\ll 1$, what reflects stronger correlation of small 
scale perturbations, and $W_s\rightarrow 1$ with increasing  
mass of pancakes.  

Analysis of simulated and observed structure at small 
redshifts (DD99; Doroshkevich, Tucker\,\&\, Allam 2002) 
shows that for richer pancakes and galaxy walls the 
cumulative and differential PDFs, $W_q\,\&\,N_q$, for the 
DM column density can be written as follows:
\be
W_q(\zeta)={\rm erf}^2(\sqrt{\zeta}),\quad N_q ={2\over
\sqrt{\pi}}e^{-\zeta}{{\rm erf}(\sqrt{\zeta})\over
\sqrt{\zeta}}{d\zeta\over dq}\,,
\label{nq1}
\ee
with the survival probability $W_s={\rm erf}(\sqrt{\zeta})$. 
However, other expressions for the survival probability can 
be also used and, for comparison, here we consider the model 
with $W_s={\rm erf}^2(\sqrt{\zeta})$ and  
\be 
W_q(\zeta)={\rm erf}^3(\sqrt{\zeta}),\quad N_q = 
{3\over\sqrt{\pi}}e^{-\zeta}{{\rm erf}^2(\sqrt{\zeta})
\over\sqrt{\zeta}}{d\zeta\over dq}\,.
\label{nq2}
\ee
This problem is further discussed in Sec. 7. 

The dimensionless mean number density of absorbers, 
$n_{abs}(q_{thr},z)$, with the above assumptions is given by:
\be
n_{abs}\approx {c\over H_0l_v}{{\rm erfc}(\zeta_{thr})
(1+z)^2\over\langle q(\zeta_{thr})\rangle}\,.
\label{nabs}
\ee
Here erfc$(\zeta_{thr})$ describes the progressive matter 
concentration within pancakes with $\zeta\geq\zeta_{thr}= 
\zeta(z,q_{thr})$ and $q_{thr}$ is the minimal (threshold) 
DM column density of pancakes in the sample under consideration 
(DD02, Paper I). For $W_s=$erf$(\sqrt{\zeta})$ and for $q_{thr}
\gg q_0$, we get  
\[
\langle q(\zeta_{thr})\rangle=4\tau^2\left[1+{4\sqrt{\pi
\zeta_{thr}}{\rm erf}(\sqrt{\zeta_{thr}})+2\exp(-\zeta_{thr})
\over \pi\exp(\zeta_{thr})[1-{\rm erf}^2(\sqrt{\zeta_{thr}}
)]}\right]
\]
where $\zeta_{thr}(z)=q_{thr}/6\tau^2(z)$. Similar relations 
can be also written for other survival probabilities. 

For pancakes with small $q_{thr}\sim q_0$ the redshift 
variations of the mean number density $n_{abs}(z,q_{thr})$ 
are described by the expression  
\be
n_{abs}\approx {c\sqrt{3}\over 4\pi H_0l_v}{(1+z)^2
\exp(-\zeta_{thr})\over \sqrt{q_0[1-\xi_v(q_{thr})]}}
W_s(\zeta_{thr})\,,
\label{nq0}
\ee
with $q_0$ as given by (\ref{qmm}) (DD02, Paper I). This 
approach considers only gaseous pancakes and, as was noted 
in Sec. 2.3, its sensitivity to the small scale perturbations 
is restricted by the Jeans damping. 

\subsection{Reconstruction of the initial power spectrum}       

Using the measured redshift, $z$, and DM column density of 
absorbers, $q$, we determine the cumulative PDF of absorbers 
$W_{obs}[>q/\tau(z)]$ and, for each $q/\tau(z)$, we compute 
$\langle q\rangle$ and $\sigma_q^2=\langle q^2\rangle-\langle 
q\rangle^2$. For a chosen $W_q(\zeta)$, we solve 
numerically the equation 
\be
W_{obs}[>q/\tau(z)]=W_q(\zeta)\,,
\label{eq1}
\ee
with respect to $\zeta(q, \tau)$ and, thus, we obtain the 
function
\be
\xi_v(q)=1-{q^2\over 4\tau^2\zeta}\,.
\label{eq11}
\ee
The observed functions $N_q(\zeta)$ and $n_{abs}(z)$ can be 
compared with corresponding expectations (\ref{nq1}) -- 
(\ref{nabs}) for the correlation function $\xi_v(q)$ 
(\ref{eq11}). This $\xi_v(q)$ can be also compared with the 
reference function $\xi_{CDM}(q)$ (\ref{xiv}).

\section{Determination of the DM column density}

To connect the main observational characteristics of 
the Lyman-$\alpha$ absorption lines, namely, the redshift, 
$z_{abs}$, the column density of neutral hydrogen, $N_{HI}$, 
and the Doppler parameter, $b$, with the dimensionless DM 
column density, $q$, we use the improved model of 
absorbers discussed in Paper I. Here we only briefly 
summarize the main assumptions and relations. Detailed 
discussion of the model can be found in Paper I. 

\subsection{Physical model of absorbers.}

Following Paper I we assume that:
\begin{enumerate}
\item{} The DM distribution forms an interconnected structure 
of sheets (Zel'dovich pancakes) and filaments. The baryonic 
component follows the DM distribution and therefore the main 
parameters of absorbers are approximately described by 
the Zel'dovich theory of gravitational instability 
(Sec. 2; DD99; DD02). 
\item{} The majority of DM pancakes are partly relaxed, 
and long-lived, but the observed properties of absorbers 
vary with time due to the successive merging, transversal 
compression and/or expansion and disruption of DM pancakes. 
\item{} Gas is trapped in the gravitational potential wells 
formed by the DM distribution. The gas temperature and the 
observed Doppler parameter, $b$, trace the depth of the DM 
potential wells. We consider the possible macroscopic 
motions within pancakes as subsonic and assume that 
they cannot essentially distort the measured Doppler 
parameter. 
\item{} The gas is ionized by the UV background and for 
the majority of absorbers ionization equilibrium is assumed.
\item{} For a given temperature, the gas density within the 
potential wells is determined by the gas entropy created 
during the previous evolution. The gas entropy is changing, 
mainly, due to shock heating in the course of pancakes 
merging, bulk heating by the UV background and local 
sources, and due to radiative cooling. Random variations 
of the intensity and spectrum of the UV background enhance 
random scatter of the observed properties of absorbers.
\end{enumerate}

This model implies that the Ly-$\alpha$ clouds trace  
the LSS in the DM component which is 
qualitatively similar to the rescaled structure observed 
at $z\leq 1$ in the galaxy distribution. It links the 
characteristics of the DM structure and the observed 
absorption lines and, as was shown in Paper I, provides 
a reasonable self consistent description, at least, of 
richer absorbers. Our model considers discrete absorbers 
and, so, it essentially differs from the popular model of 
"fluctuating Gunn--Peterson approximation" discussed more thoroughly
in Sec. 9.4.   

Our approach cannot be applied to objects for which the 
gravitational potential and the gas temperature along the line 
of sight strongly depends on the matter distribution across this 
line. However, the anisotropic halos of filaments and clouds with 
moderate $N_{HI}$ can be considered as 'pancake-like'.

Special attention require description of weaker absorbers with 
$b\leq b_{bg}$. This subpopulation contains "artificial" caustics 
(McGill 1990) and absorbers identified with slowly expanding 
underdense regions (Bi \& Davidsen 1997; Zhang et al. 1998; 
Dav\'e et al. 1999). Such absorbers produce a short lived noise, 
which is stronger at higher redshifts $z\geq$~3. However, such 
absorbers can also be formed within low temperature regions
arising owing to spatial nonhomogeneities of the UV background 
and expansion rate. Strictly speaking our model does not 
include such absorbers because we assume that $b\geq b_{bg}$ 
for all absorbers formed due to compression of background 
matter and we suppose that $b_{bg}$ is constant and it does 
not depend on the position or time. However, we can include 
absorbers with $b\leq b_{bg}$ into consideration assuming that 
they are adiabatically compressed but not bound gravitationally. 

\subsection{DM column density of absorbers}

In this section we use the physical model discussed in 
Sec. 3.1 to introduce the basic relations between 
theoretical and observed characteristics of pancakes. 
The most 
important characteristics of DM pancakes are presented 
(without proofs) as a basis for further analysis. More 
details are given in DD99, DD02, and Paper I.

The fundamental characteristic of DM pancakes is the 
dimensional, $\mu(z)$, or the dimensionless, $q$, DM 
surface density (the dimensionless DM column density) :
\be
\mu(z)\approx {\langle\rho_m(z)\rangle l_vq\over 
(1+z)}\,, 
\label{mu}
\ee
where the coherent length of the initial velocity field $l_v$ 
was introduced in (\ref{lv}). As was discussed in Sec. 2.4, 
$l_v q$ is defined as the unperturbed distance at redshift 
$z=0$ between DM particles bounding the pancake. 

As was assumed in Sec. 3.1, we consider majority of absorbers as
gravitationally bound and relaxed long lived DM pancakes.  Their
properties are slowly changing mainly owing to the process of merging
and transverse expansion and compression. For such absorbers, the gas
temperature, $T_g$, and the Doppler parameter, $b\propto\sqrt{T_g}$
(\ref{bg}) are closely linked with the gravitational potential of the
DM pancake and we get: 
\be
\beta^2={b^2\over b_{bg}^2}\approx 1+{4\over 5}
{\pi G\mu^2\Theta_\Phi\over b_{bg}^2\langle \rho(z)
\rangle\delta_m}=1+\delta_0{q^2\over\delta_m}(1+z)\,,
\label{bb}
\ee
\[
\delta_0 = {3\over 10}\left({H_0l_v\over b_{bg}}\right)^2
\Omega_m\Theta_\Phi\approx 4\cdot 10^3
{0.13\over\Omega_m h^2}
\left({16{\rm km/s}\over b_{bg}(z)}\right)^2\Theta_\Phi.
\]
Here $\delta_m$ is the mean overdensity of the DM pancake 
and the random factor $\Theta_\Phi\sim 1$ describes the 
nonhomogeneity of DM distribution across the pancake. 

Under the condition of ionization equilibrium,  for the 
column density of neutral hydrogen, we get
\be
N_{HI}={2\cdot 10^{16}{\rm cm}^{-2}\over \delta_0G_{12}(z)}
\delta_bq\beta^{-3/2}(1+z)^5
\label{NH2}     
\ee
\[
= {2\cdot 10^{16}{\rm cm}^{-2}\over G_{12}(z)}
{q^3(1+z)^6\over\beta^{7/2}}{\beta^2\over\beta^2-1}
{\delta_b\over\delta_m}\,,
\]
\be
G_{12}={\Gamma_{12}(z)\over\Theta_g(z)}{\cos\theta\over 
0.5}\left({\Omega_mh^2\over 0.13}{0.02\over\Omega_bh^2}
\right)^2\left({b_{bg}\over 16{\rm km/s}}\right)^{7/2}
\label{gg}
\ee
where $\Gamma_{12}=\Gamma_\gamma/10^{-12}s^{-1}$, 
$\Gamma_\gamma$ characterizes the rate of ionization 
of hydrogen by the UV background, $\delta_b$ is the 
overdensity of baryonic component, cos$\theta$ 
describes a random orientation of the absorber and the 
line of sight and the function $\Theta_g(z)\sim 1$ 
describes the action of several random and unknown 
factors (more details are given in Paper I). 

The equations (\ref{bb}) and (\ref{NH2}) relate three 
independent variables, namely, $q, \delta_b\,\&\,\delta_m$. 
To find the DM column density, $q$, it is therefore necessary 
to use an additional relation which connects the basic 
properties of absorbers. In Paper I we assumed that $\beta
\geq 1$, $\delta_m\approx\delta_b$. These assumptions are 
valid for richer absorbers formed due to strong shock 
compression when the essential growth of the entropy and 
temperature suppresses differences between the DM and 
baryonic components and so their distribution across the 
absorber is similar. For the DM column density of such 
absorbers we get 
\be
q^3\approx\left({N_{HI}\over 2\cdot 10^{16}{\rm cm}^{-2}}\right)
{\beta^{3/2}(\beta^2-1)\over (1+z)^6}G_{12}(z)\,.
\label{qq}
\ee
This relation describes correctly properties of hot absorbers 
but decreases the column density of rare absorbers with 
$\delta_m\leq\delta_b$. 

However, formation of weaker DM absorbers with $\beta 
\sim 1$ is accompanied by adiabatic inflow of baryons 
into the potential well. In this case, in accordance 
with the polytropic index $\gamma_b=5/3$ for baryons,
we can expect that 
\be
\delta_b\approx\beta^3,\quad \delta_m = \delta_0{q^2(1+z)
\over\beta^2-1}\geq\delta_b\,,
\label{adlim}
\ee 
\[
q\geq \beta^{3/2}(\beta^2-1)^{1/2}\delta_0^{-1/2}
(1+z)^{-1/2}\,,
\]
where $\delta_0$ was introduced in (\ref{bb}).
This expression quantifies the bias between 
the compressed DM and baryonic components arising due to 
the impact of gaseous pressure (Jeans damping). For such 
absorbers we get from (\ref{NH2}): 
\be
q\approx\left({N_{HI}\over 2\cdot 10^{16}{\rm cm}^{-2}}\right)
{G_{12}\delta_0\over\beta^{3/2}(1+z)^{5}}\,,
\label{qad}
\ee
 
Both expressions, (\ref{qq}) and (\ref{qad}), define the 
dimensionless column density of DM component corrected 
for the impact of gaseous pressure. These relations can 
be successfully applied to absorbers formed due to 
adiabatic and strong shock compressions. However, 
the boundary between these limiting cases must 
be established {\it a priori}. So, to discriminate 
absorbers described by (\ref{qq}) and (\ref{qad}), 
we can use a threshold Doppler parameter, $\beta_{thr}$,
which, in fact, characterizes the Mach number of the 
inflawing matter. Thus, the weak adiabatically compressed 
absorbers can be conveniently separated with a threshold, 
\be
1\leq \beta\leq \beta_{thr}\approx 1.5 - 2\,, 
\label{betalim}
\ee
under condition (\ref{adlim}) which can be reformulated as 
a restriction on the column density of neutral hydrogen, 
$N_{HI}$. Absorbers with $\beta\geq\beta_{thr}$ are 
considered as dominated by shock compressed gas and their 
DM column density is given by (\ref{qq}). 

As was noted above, we cannot correctly determine 
properties of absorbers with $b\leq b_{bg}$ in the 
framework of our model because they violate the relation 
(\ref{bb}). However, we can consider them 
as adiabatically compressed, short lived unrelaxed and 
gravitationally unbounded pancakes. Under these conditions
absorbers with $b\leq b_{bg}$ can be described by the expression 
(\ref{qad}) which does not depend on $b_{bg}$.

As was found in Paper I, we can approximate the combined 
action of redshift variations of the UV background and 
random factors in expressions (\ref{qq}\,\&\,\ref{qad}) 
by a two parameters relation 
\be
\langle G_{12}(z)\rangle\approx G_0(1+\alpha_\gamma z)\,,
\label{g12}
\ee
where $G_{0}$ and $\alpha_{\gamma}$ are constants.
The precision of this approach is moderate but a 
suitable choice of $G_0$ and $\alpha_\gamma$ provides 
a reasonable statistical description for majority of 
absorbers. 

\subsection{Statistical characteristics of absorbers}

The model of absorbers discussed in this section is  
in fact independent from the statistical 
characteristics discussed in Sec. 2\,. It introduces the 
physically motivated combination of observed absorber 
characteristics, $q$, as given by (\ref{qq}) and (\ref{qad}), which 
can be compared with theoretical expectations introduced 
in Sec. 2\,. As it is shown in Sec. 5, the functions $\zeta(q,z)$ 
(\ref{xit}) and $W_q$ found with this approach are  
consistent with expectations what verifies the 
interpretation of $q$ as the DM column density of 
absorbers. 

\section{Mean characteristics of the observed sample}

\subsection{The database}

In our analysis we use 14 high resolution QSO spectra 
listed in Table I\,. To decrease the scatter, 
we selected the sample of absorbers requiring that 
$10^{12}{\rm cm}^{-2}\leq N_{HI}\leq 10^{15}{\rm cm}^{-2}$, 
$b\leq$ 90 km/s. 

Absorbers with $b\leq b_{bg}$ must be excluded from the sample because
for them the overdensity $\delta_b=(b/b_{bg})^{2/3}\leq$ 1\,. For
$b_{bg}=16$km/s our sample includes 3752 absorbers at redshifts 1.7
$\leq z\leq$ 4\,.  Subpopulation of $\sim 500$ absorbers with $b\leq
b_{bg}$ contains "artificial" caustics (McGill 1990) and absorbers
which are most probably situated within regions where the average
density and temperature are smaller than the values specified in
(\ref{bg}) (Bi \& Davidsen 1997; Zhang et al.  1998; Dav\'e et
al. 1999). As was noted above we can include them into consideration
assuming that they are short lived gravitationally unbound pancakes.

For all models discussed in this paper we describe 
the redshift variations of the UV background intensity 
by the relation (\ref{g12}) with  
\be
G_0=12 - 15,\quad \alpha_\gamma=-0.22\,.
\label{uv}
\ee
Estimates of the intensity of UV background through the 
proximity effect (Scott et al. 2000) give $\Gamma_{12}
\approx 2\pm 1$ at $z\sim 3$. Similar estimates were 
obtained in McDonald \,\&\,Miralda-Escude (2001). These 
results are consistent with (\ref{gg}), (\ref{g12}) 
and (\ref{uv}) when 
\[
\langle\Theta_g\rangle\sim 0.4(b_{bg}/16{\rm km/s})^{-7/2}\,.
\]
Such a choice is in accordance with expectations of our
model of absorbers and results in weak redshift variations 
of the function $\langle\zeta(z)\rangle$. 

\subsection{Redshift variations of absorbers characteristics}

\begin{figure}
\plotone{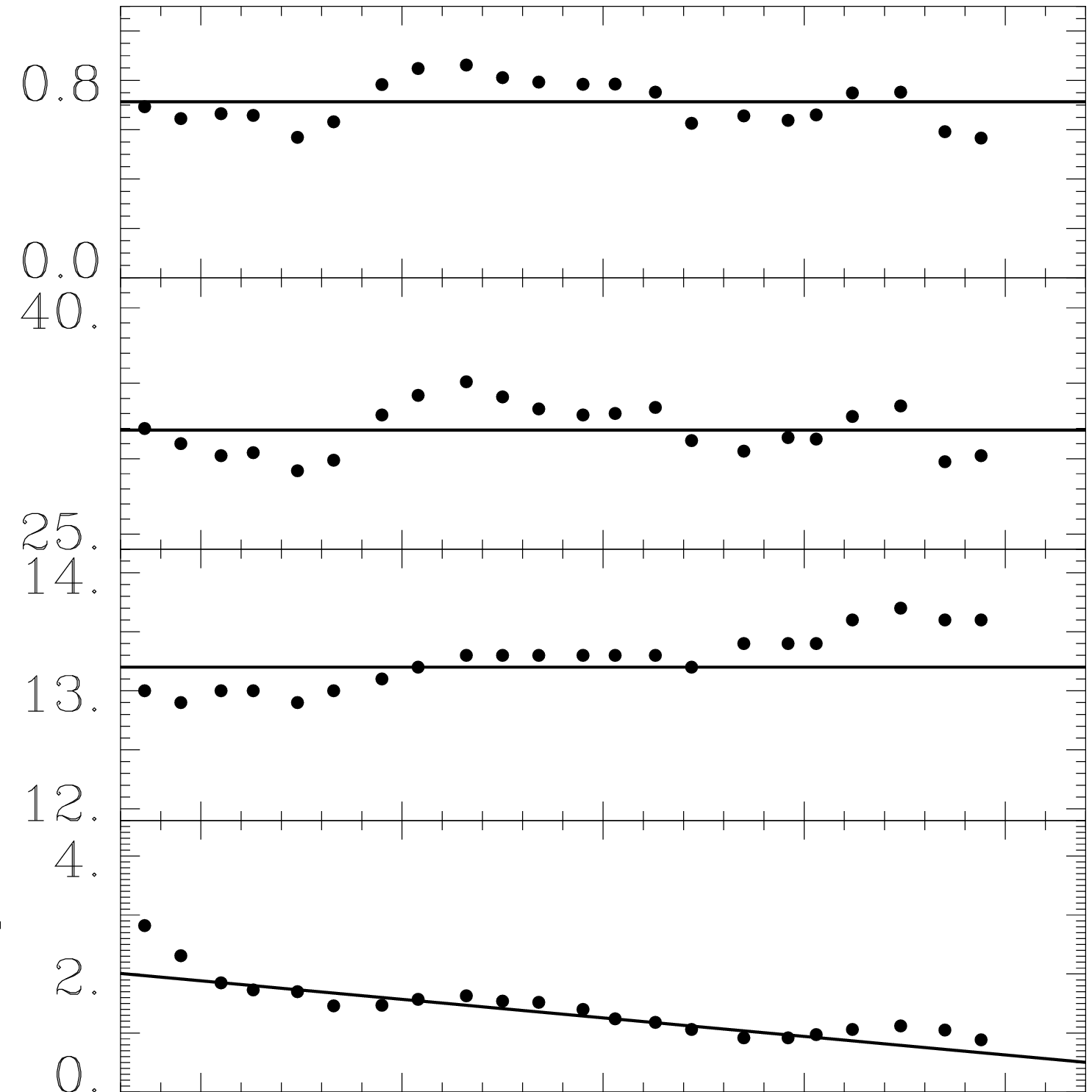}
\vspace{1.4cm}
\caption{The mean DM column density of absorbers, 
$\langle\zeta\rangle$, the mean Doppler--parameter, 
$\langle b\rangle$ (in km/s), the mean neutral hydrogen column 
density, $\langle {\rm log}N_{HI}\rangle$ and the mean absorber 
separations, $\langle D_{sep}\rangle h^{-1}$Mpc, are plotted 
vs. redshift $z$. Fits (\ref{zmns}) are plotted by 
solid lines.
} 
\label{fig1}
\end{figure}

Here we consider the redshift variations of the observed 
parameters of absorbers, namely, the neutral hydrogen column 
density, ${\rm log}N_{HI}$, the Doppler parameter, $b$, and 
distances between absorbers
\be
D_{sep}={c\Delta z_i\over H(z)}\approx 5.5\cdot 
10^3{\Delta z_i\over (1+z)^{3/2}}\sqrt{0.3\over\Omega_m}
h^{-1}{\rm Mpc}\,.
\label{sep}
\ee
For our sample of absorbers, the redshift variations of 
the mean observed parameters and the mean DM column density 
measured by the self similar variable, $\langle\zeta(z)
\rangle$ (\ref{xit}), are plotted in Fig. \ref{fig1} together with 
approximate fits :
\be
\langle\zeta(z)\rangle\approx 0.73\pm 0.08,\quad 
\langle b(z)\rangle\approx (31.9\pm 1.6){\rm km/s}\,,
\label{zmns}
\ee
 \[
\langle {\rm log}N_{HI}\rangle\approx 13.2\pm 0.22,\quad 
\langle D_{sep}\rangle\approx (2.5-0.6 z)h^{-1}{\rm Mpc}\,.
\]
These results are consistent with the main results obtained 
in Paper I. It is interesting to notice that the redshift 
variations of $\langle\zeta(z)\rangle$ and $\langle b(z)
\rangle$ plotted in Fig. \ref{fig1} are clearly correlated.  

\begin{figure}
\plotone{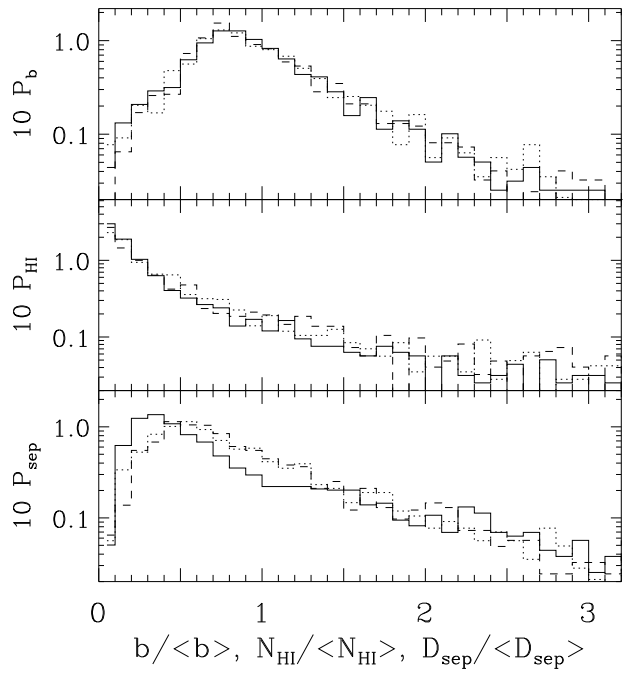}
\vspace{1.6cm}
\caption{PDFs for the Doppler parameter, $P_b$, the 
hydrogen column density, $P_{HI}$, and absorbers 
separation, $P_{sep}$, for redshift 1.7 -- 2.5 (solid lines), 
2.5 -- 3 (dot lines) and 3 -- 4.5 (dashed lines).  
} 
\label{fig2}
\end{figure}

The PDFs $P_b$, $P_{HI}$ and $P_{sep}$, for the Doppler 
parameter, $b$, the hydrogen column density, $N_{HI}$, and 
the absorbers separation, $D_{sep}$, are plotted in Fig. 
\ref{fig2} for three redshift intervals, namely, 1.7 -- 2.5 
(1357 lines), 2.5 -- 3 (1235 lines) and 3 -- 4.5 (1160
lines). This Fig. indicates that the shapes of PDFs weakly vary
with redshift, what is consistent with the 
assumption that  absorbers are long--lived and relaxed.

The mean values of the measured $\langle b\rangle$ and 
$\langle N_{HI}\rangle$ for these intervals of $z$ are quite 
stable:
\[
\langle b\rangle = 28.3{\rm km/s},\quad 31.1{\rm km/s},
\quad 30.0{\rm km/s}\,,
\]
\be
\langle N_{HI}\rangle = 10^{13.5}{\rm cm}^{-2},\quad 10^{13.6}
{\rm cm}^{-2},\quad 10^{13.8}{\rm cm}^{-2}\,,
\label{tmns}
\ee
\[
\langle D_{sep}\rangle = 1.75 h^{-1}{\rm Mpc},\quad 1.34 
h^{-1}{\rm Mpc},\quad 0.94 h^{-1}{\rm Mpc}\,.
\]

To quantify the redshift variations of the PDFs we calculate 
the mean values and dispersions of the relative differences, 
\[
c_i^k=[P_i^{(t)}-P_i^k]/[P_i^{(t)}+P_i^k]\,,
\] 
between PDFs for the full sample ($P_i^{(t)}$, 3752 lines) 
and for each individual redshift interval ($P_i^k$, k=1, 
2, 3), where $i=1, 50$ numbers the bins with a step of 0.1 
of the mean value. For the most representative first 25 bins 
$\langle c_i^k\rangle=0$ with the standard deviation 
$\sigma_{c}\sim 0.1 - 0.15$. 

These results indicate also the homogeneity of our sample.
They confirm that the redshift evolution of the observed  
parameters of absorbers is very similar to the evolution 
of their mean values (\ref{zmns}). It is an evidence in 
favor of the model of long--lived gravitationally confined 
absorbers. More detailed discussion of the observed spectra can 
be found in Kim, Cristiani \& D'Odorico 
(2002) and Kim et al. (2002). 

It is interesting to estimate also the mean matter fraction 
accumulated by absorbers, $\langle f_m\rangle$. Rough estimates 
of $f_m$ can be found by comparing the unperturbed mean size, 
$l_v\langle q\rangle$, with the mean separation of absorbers, 
$D/N_{abs}$ for each individual spectrum, we get
\be
\langle f_m\rangle\approx N_{abs}l_v\langle q\rangle/D\approx 
0.87\pm 0.14\,. 
\label{fraction}
\ee
Here $N_{abs}$ and $D$ are the number of absorbers in the  
spectrum and its length along the line of sight. Averaging is 
performed over all fourteen spectra in our data base.

This value $\langle f_m\rangle$ exceeds the theoretical expectation 
$\langle f_m\rangle\sim 0.5 - 0.6$ at $z\sim 3$ (Paper I) what 
indicates a limited precision of this approach. In spite of this, 
it verifies that, in accordance with the physical model, at redshifts 
$z\leq 3$ at least half of matter could be accumulated by absorbers. 
This estimate does not contradict the observed  
Gunn--Peterson effect in HeII at $z\sim 3$  (Jacobsen et al. 1994) 
what indicates that essential fraction of baryons is homogeneously
distributed. However quantitative estimates of this fraction strongly 
depends upon the intensity of UV background ionizing HeII.

\section{Correlation function of the initial velocity field}

In this Section we consider the cumulative and differential 
distribution functions, $W_q(\zeta)$ and $N_q(\zeta)$, and 
the correlation function of the initial velocity field, 
$\xi_v$, which can be found from $W_q(\zeta)$ as was described 
in Sec. 2.5\,. These functions depend upon several parameters 
which are known with limited precision. To test their influence 
we consider below three models of absorbers.

\subsection{Three models of absorbers}

Our results weakly depend upon the parameter $\beta_{thr}$ 
introduced in Sec. 3.2 to separate absorbers formed by adiabatic 
and shock compression. The choice of $\beta_{thr}$ is restricted 
by the shape of the function $\xi_v$ at larger $q$ (\ref{xiv}) as
$\beta_{thr}\leq 2$. In our model the amplitude of initial
perturbations and
\be
\tau_0=0.22,\quad \beta_{thr}=1.5,\quad b_{bg}=16 km/s\,,
\label{taub}
\ee
remain fixed.
As is seen from (\ref{tmns}), $\beta_{thr}\sim\langle b\rangle
/b_{bg}$ and it turns out that $\sim 700-1\,000$ of absorbers 
in our sample have been adiabatically compressed. They form the 
left side of the PDF $P_b$ plotted in Fig. \ref{fig2} while richer 
absorbers with $\beta\geq \beta_{thr}\sim\langle b\rangle/b_{bg}$ 
have been presumably formed  by  merging and shock compression.

The cumulative and differential distribution functions depend 
also upon the survival probability, $W_s=$erf$^n(\sqrt{\zeta})$, 
(\ref{nq1}, \ref{nq2}) described by the power index $n$. Finally, 
of course our results depend upon the 
number of absorbers in the sample, $N_{smp}$, used in the analysis. 
In particular, in 
spite of the relatively small number of absorbers with $b\leq 
b_{bg}$ when they are included they noticeably change
the correlation function $\xi_v$. 

To test the influence of these factors we consider three models 
with parameters:
\be
n=1,\quad N_{smp}=3\,752,\quad G_0=12.5\,,
\label{m400}
\ee
\be
n=1,\quad N_{smp}=4\,248,\quad G_0=12.5\,,
\label{m402}
\ee
\be
n=2,\quad N_{smp}=4\,248,\quad G_0=15.5\,,
\label{m403}
\ee
for the Model 1, Model 2 and Model 3, respectively. Parameter 
$G_0$ was introduced in (\ref{g12}) to characterize the intensity 
of the UV background. The Model 1 includes only absorbers with 
$b\geq b_{bg}$ whereas Models 2 \& 3 include also absorbers with 
$b\leq b_{bg}$ and differ by the choice of survival probability, 
$W_s$, and the intensity of the UV background. 

\subsection{Cumulative PDFs for dimensionless DM column 
density}

As was discussed in Sec. 2.5, the observed cumulative 
distribution function, $W_q(>q/\tau)$, can be estimated 
using $q/\tau$ obtained from our sample of observed 
absorbers with the model described in Sec. 3 and 
parameters (\ref{taub}-\ref{m403}). For these models 
the correlation function $\xi_v$ is well fitted by the 
expression
\be
\xi_{fit}=1-1.5q\sqrt{q^2+p_1^2\over q^2+p_2^2},\quad 
0.5\geq q\geq 10^{-3}\,,
\label{xi400}
\ee
\[
p_1=1.3\cdot 10^{-2},\quad p_1= 10^{-2},\quad p_1=2\cdot 10^{-3}\,,
\]
\[
p_2=7\cdot 10^{-4},\quad p_2=4\cdot 10^{-3},\quad p_2= 2\cdot 10^{-3}\,,
\]
for the Model 1, Model 2, and Model 3, respectively. The function 
(\ref{xi400}) becomes identical with the reference correlation 
function (\ref{xiv}) $\xi_{CDM}$ for $q\geq p_1\geq p_2$. 

\begin{figure}
\plotone{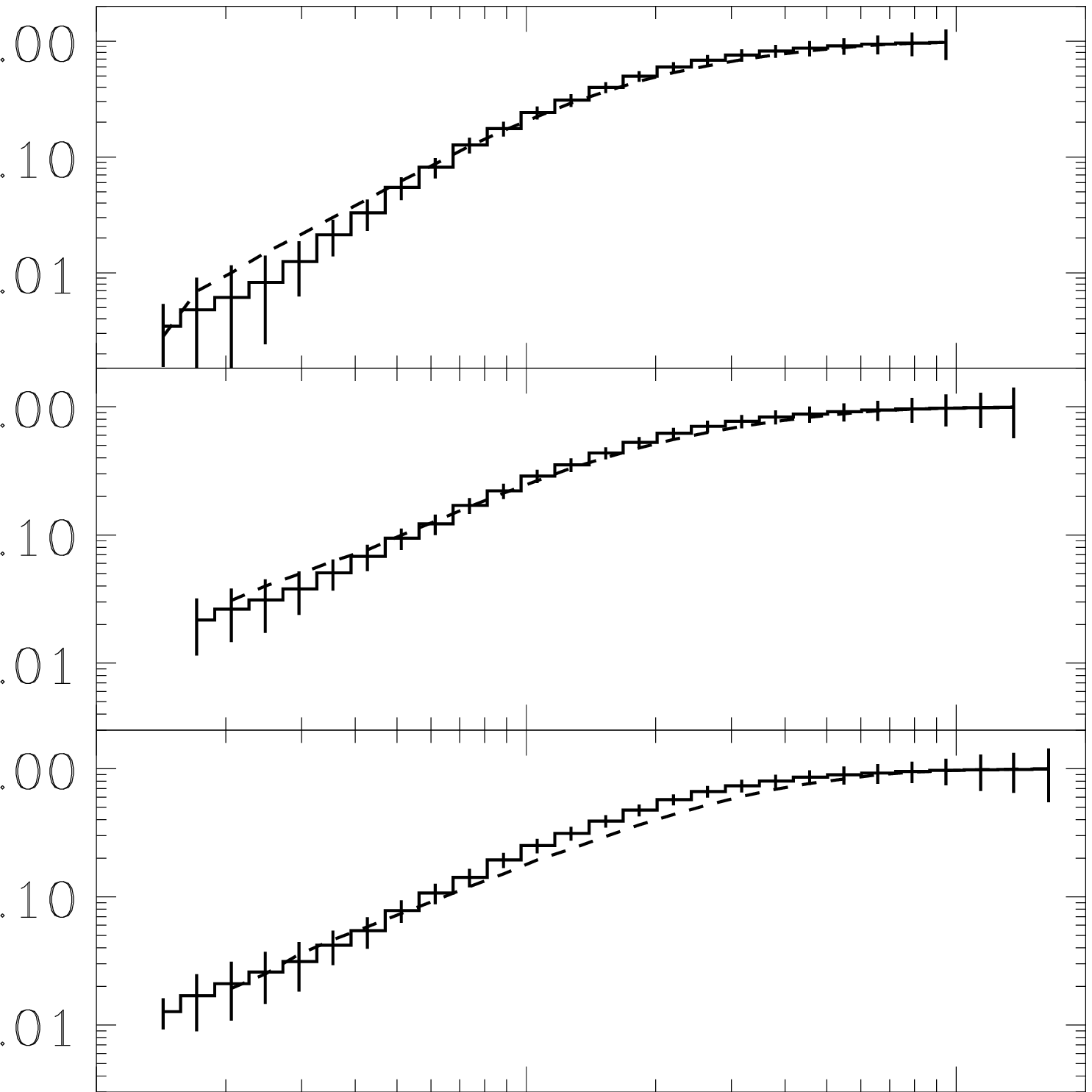}
\vspace{1.5cm}
\caption{Cumulative distribution functions, $W_q$, are 
plotted vs. $q/\tau$ for Model 1 (top panel), Model 2 
(middle panel) and Model 3 (bottom panel). In the same 
panels, the best fits (\ref{xi400})  
are plotted by solid lines. 
} 
\label{fig3}
\end{figure}

The measured functions $W_q$ for the three models are plotted 
in Fig. \ref{fig3} together with fits (\ref{nq1}) and 
(\ref{nq2}) for $\xi_v$ given by (\ref{xi400}) and with a 
scatter $\propto 2/\sqrt{N_{abs}}$ where $N_{abs}$ is the 
number of absorbers in each bin. The difference between the 
measured and fitted functions $W_q$ is characterized by the 
relations:
\be
W_q\approx \left\{ \begin{array}{ll}
 (0.96\pm 0.2){\rm erf}^2(\sqrt{\zeta}), & \chi^2 = 0.5\,,\cr
 (1.0\pm 0.1){\rm erf}^2(\sqrt{\zeta}), & \chi^2 = 0.2\,,\cr  
 (1.1\pm 0.2){\rm erf}^3(\sqrt{\zeta}), & \chi^2 = 0.5\,,\cr
  \end{array}  \right.
\label{wfit}
\ee
where $\zeta$ is given by (\ref{xit}) and (\ref{xi400}). 
$\chi^2$ is found for ratios of measured and 
fitted $W_q/$erf$^{n+1}(\sqrt{\zeta})$ for 26 measured 
points. For $\xi_v$ given by (\ref{xi400}) all fits are 
reasonably consistent with the measured $W_q$. 

\subsection{Differential PDFs for dimensionless DM column 
density}

For all three models the differential PDFs were found 
from measured $q$ and $\tau$ with $\xi_v$ given by 
(\ref{xi400}). These PDFs are quite sensitive to the 
functions $\xi_v$ and $W_s$, what allows one to check 
the correctness of the choice of these functions.

\begin{figure}
\plotone{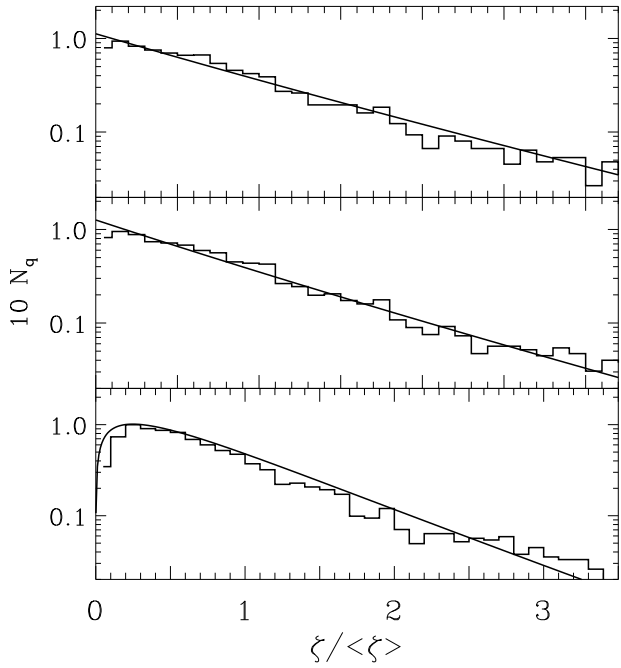}
\vspace{1.5cm}
\caption{PDFs, $N_q$, for the Model 1 (top panel),
the Model 2 (middle panel) and the Model 3 (bottom panel) 
are plotted vs. $\zeta/\langle\zeta\rangle$. Solid lines 
shows fits (\ref{nq1}) and (\ref{nq2}) for $\xi_v$ given 
by (\ref{xi400}). 
} 
\label{fig4}
\end{figure}

The differential PDF, $N_q(\zeta)=\langle\zeta\rangle 
dW_q/d\zeta$, is plotted in Fig. \ref{fig4} for the Model 
1, Model 2, and Model 3 together with the fit 
\be
N_q(\zeta)\approx {(n+1)\langle\zeta\rangle\over
\sqrt{\pi}}\exp(-\zeta){{\rm erf}^n(\sqrt{\zeta})\over
\sqrt{\zeta}}\,.
\label{pdf1}
\ee
This expression fits well the observed PDF for 
\be
\langle\zeta_{th}\rangle\approx\langle\zeta\rangle=0.85,
\quad 0.87,\quad 1.15\,,
\label{zetafit}
\ee
for Models 1, 2, \& 3, respectively. Here $\langle\zeta_{th}
\rangle=0.82\,\&\,1.05$ are the mean values for $N_q$ (\ref{pdf1}) 
with $n=1$ and $n=2$. For $\zeta\leq 0.2\langle\zeta\rangle$ the 
measured PDF becomes sensitive to the possible incompleteness of 
the sample. 

The concordance of  the measured and expected PDFs 
can be also checked by the Kolmogorov--Smirnov test. For 
the three models plotted in Fig. \ref{fig4} the cumulative 
PDFs are consistent with erf$^{n+1}(\sqrt{\zeta})$ with 
probabilities 
\be
P_{KS}=0.78,\quad P_{KS}=0.76,\quad P_{KS}=0.47\,,
\label{ks}
\ee
respectively. These results confirm, for a given set of 
DM column density, a close link between the accepted survival 
probability, $W_s$, and the measured correlation function, 
$\xi_v$. For $\zeta\geq\langle\zeta\rangle$ the exponential 
shape of the function $N_q$ coincides with its theoretically 
expected shape  for Gaussian initial perturbations and the CDM--like 
power spectrum (\ref{xiv}), (\ref{xit}), (\ref{nq1}). 

\subsection{Correlation function of the initial 
velocity field}

As was described in Sec. 2.5 we can derive the 
correlation function of the initial velocity field, 
$\xi_v(q)$, by solving Eq. (\ref{eq1}). This function 
depends on both the survival probability, $W_s$, and 
the sample of absorbers (see \ref{taub}-\ref{m403}).

\begin{figure}
\plotone{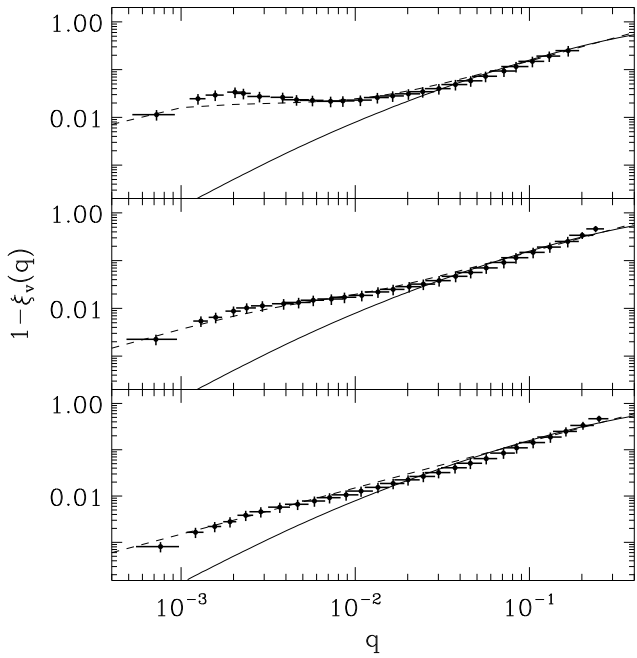}
\vspace{1.5cm}
\caption{Correlation function of the initial 
velocity field, $\xi_v(q)$, for the Model 1 (top panel), 
Model 2 (middle panel) and Model 3 (bottom panel). 
Dashed and solid lines show fits (\ref{xi400}) and 
the function $\xi_{CDM}$ (\ref{xiv}).
} 
\label{fig5}
\end{figure}

For all three models, the measured correlation functions 
$\xi_v(q)$ are plotted in Fig. \ref{fig5} together with 
$\xi_{fit}$ (\ref{xi400}) and with the correlation function 
$\xi_{CDM}$ (\ref{xiv}). For all models variations of 
the ratios of measured and fitted functions, $(1-\xi_v)
/(1-\xi_{fit})$, are quite moderate and are quantified 
as follows: 
\be
(1-\xi_v)/(1-\xi_{fit})\approx 1\pm 0.2,\quad 0.3\geq q\geq 1
0^{-3}\,,
\label{xifit}
\ee
\[
\chi^2 = 0.9,\quad 0.4,\quad 0.7\,, 
\]
for the  Model 1, Model 2 and Model 3, respectively.
Here $\chi^2$ are found from 26 measured points. This 
fact confirms the self consistency of our models and 
the choice of parameters (\ref{uv}) and (\ref{taub}-\ref{m403}). 

At the same time, at larger scales, $0.3\geq q\geq 0.03,\,
10h^{-1}$Mpc$\geq l_vq\geq 1 h^{-1}$Mpc, for all three Models 
the differences between the measured and CDM--like correlation 
functions are negligible
\be
(1-\xi_v)/(1-\xi_{CDM})\approx 1\pm 0.15\,,
\quad \chi^2 = 0.2\,,
\label{xi11}
\ee
where $\chi^2$ are found for the ratios $(1-\xi_v)/(1-\xi_{CDM})$
for 14 measured points. These results demonstrate that for 
such scales the differences between the CDM--like initial 
power spectrum and the measured one do not exceed 15\%. 

However, at small scales all measured correlation functions 
differ from the $\xi_{CDM}$ (\ref{xiv}). Quantitatively these 
differences are characterized by the $\chi^2_{CDM}$ found 
for the ratios $(1-\xi_v)/(1-\xi_{CDM})$ for 26 measured points. 
For the Model 1, Model 2 and Model 3 we get, respectively :
\be
\chi^2_{CDM} = 550,\quad \chi^2_{CDM} = 120,\quad 
\chi^2_{CDM} = 33\,.
\label{xicdm}
\ee
These results show that this difference between the measured 
$\xi_{v}$ and $\xi_{CDM}$ decreases when 
formation of pancakes with small mass is suppressed more
strongly i.e. when the survival probability is 
$W_s=$erf$^2(\sqrt{\zeta})$.

\section{Redshift distribution of absorbers}

The redshift distribution of absorbers only weakly depends  
on the properties of weaker absorbers because 
they are relatively rare, only about $\sim$ 20 \% of 
absorbers have $q\leq$ 0.03. It depends mainly upon measured 
redshifts of absorbers and is only weakly sensitive to the 
physical model of absorbers discussed in Sec. 3. This means 
that it is an important independent test of the model of 
absorbers formation considered in Sec. 2.

\begin{figure}
\plotone{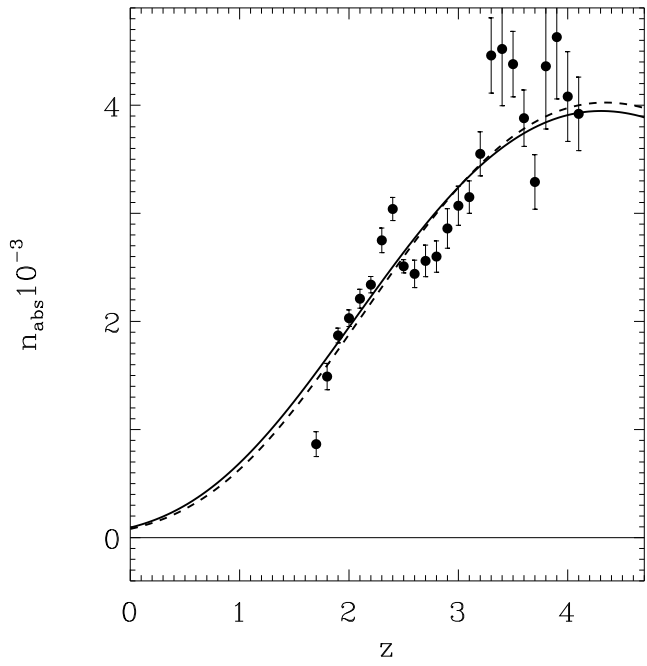}
\vspace{1.5cm}
\caption{Redshift distribution of absorbers, $\langle 
n_{abs}(z)\rangle$, for 4500 observed absorbers. Fits 
(\ref{nabs}) for Model 1 and Model 3 are plotted by  
solid and dashed lines.
} 
\label{fig6}
\end{figure}

The redshift distribution of absorbers 
is quite well fitted 
by the relation (\ref{nabs}) which is applied to all samples 
of absorbers with $q_{thr}\geq q_0$. To illustrate this 
property we plot in Fig. \ref{fig6} the measured function 
$\langle n_{abs}(z)\rangle$ for all 4\,250 absorbers together 
with two fits (\ref{nabs}). 

As was described in Paper I, for each redshift interval, $\Delta
z=0.1$, all absorbers were organized into an "equivalent single field"
by arranging them one after the other along the line of sight. The
mean number density of absorbers, $n_{abs}$, and its scatter were
found by comparing the equivalent single field with a corresponding Poissonian
distribution.  The measured function $n_{abs}$ is plotted in
Fig. \ref{fig6} together with the best fits (\ref{nabs}) for Model 1
and Model 3. Both fits are very similar to each other what
demonstrates that the observed redshift distribution of absorbers only
weakly depends on the shape of the initial power spectrum.  However,
the observed redshift distribution of absorbers verifies the
Gaussianity of the initial perturbations since it is manifested in the
characteristics of the Ly-$\alpha$ forest.

\section{Reliability and precision of our estimates}

To measure the initial power spectrum it is 
necessary to use a model of absorbers which links the observed 
Ly-$\alpha$ clouds with the DM spatial distribution. The model 
introduced in Secs. 2\,\&\,3 provides a quite reasonable and 
self consistent statistical description of the basic properties 
of absorbers (Paper I and Secs. 4\,\&\,5). For the three models 
under consideration the measured correlation function of the 
initial velocity field, $\xi_v$, is consistent with the CDM--like 
one for $q\geq 0.03$, what verifies the reliability of our approach. 

However, at small scales ($q\leq 0.03$) the measured function 
$\xi_v$ differs from the CDM--like one. These differences are caused 
by the deficit of absorbers with small $q/\tau\,\&\,q$ and, therefore, 
it can be artificially enhanced by an action of several factors. The 
most important ones are the possible incompleteness of the observed 
sample of weaker absorbers, overestimates of the DM column density 
by our model of absorbers (\ref{qq}, \ref{qad}), and the choice of 
the survival probability (\ref{nq1}, \ref{nq2}). 

Usual estimates of the possible incompleteness of the 
sample ($\leq 20 \%$) are based on the quality of the 
observed spectra but for compiled samples of absorbers 
such estimates are not reliable. The best way to test the 
influence of this factor is to repeat the analysis with 
a richer sample of observed absorbers. 

By comparing  the Model 1 with the Model 2 we  test 
the sensitivity of our results to variations of the observed 
sample. In spite of the fact that the condition $b\leq b_{bg}$ 
is satisfied only by a relatively small fraction of absorbers 
($\approx$ 500) their influence is quite significant 
because when they are included they almost double the number 
of weaker absorbers. 
In particular, the difference between the measured $\xi_v$ and 
the reference one (\ref{xiv}) is reduced by half but it is not 
eliminated. As is seen from (\ref{xicdm}), this difference 
remains clearly noticeable even for the Model 3 for which both factors, 
larger number of absorbers and larger power index, $n=2$, 
of the survival probability work together to decrease this difference.  
 
On the other hand, extension of the sample by adding 355 
absorbers with $10^{13}{\rm cm}^{-2}\leq N_{HI}\leq 10^{15}
{\rm cm}^{-2}$ from published spectra of four quasars 
(2126-158, Giallongo et al. 1993; 1700+642, Rodriges et al. 
1995; 1225+317, Khare et al. 1997; 1331+170, Kulkarni et al. 
1996)  does not change our previous results (\ref{xifit}, 
\ref{xicdm}). This test demonstrates that the measured 
velocity correlation function strongly depends on the 
representativity of the sample of weak absorbers and 
estimates of their Doppler parameter but it only weakly 
depends on the sample of stronger absorbers. 

Our choice of the survival probability in the expression 
(\ref{nq1}) is based on the measured distribution of walls 
in observed galaxy surveys and in simulations rather than on 
theoretical arguments. Comparison of results obtained with 
the expressions (\ref{nq1}\,\&\,\ref{nq2}) illustrates the 
importance of this factor and shows that the measured velocity 
correlation function could be noticeably different from that 
derived in a standard way from the CDM--like power spectrum 
only for a moderate survival probability, $W_s(q,\zeta)$. The 
analysis repeated for a wider set of functions $W_s(q,\zeta)$ 
confirms that the measured functions $W_q$ and $N_q$ strongly 
restrict the acceptable shape of the survival probability, but 
for a suitable choice of $W_s$, the difference between 
$\xi_{CDM}$ and the measured velocity correlation function 
becomes moderate. 

\begin{figure}
\plotone{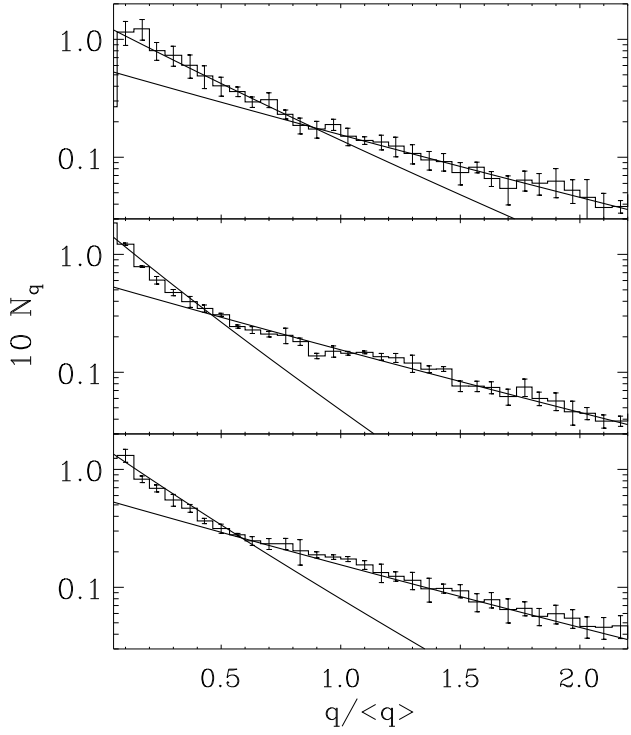}
\vspace{1.5cm}
\caption{The PDFs of dimensionless surface density of 
walls, $N_q$, for 3 sets of walls selected from two high resolution 
DM simulations with small threshold richness. 
Fits (\ref{nq1}) are plotted by solid lines.
} 
\label{fig7}
\end{figure}

These results show that the merging of low mass pancakes 
is an important factor and it should be thoroughly investigated. 
As the first step of such investigation the PDFs $N_q$ were 
measured for three sets of DM walls selected at $z=0$ with a 
small threshold richness from two standard high resolution 
simulations (Cole et al. 1998; Jenkins et al. 1998). These 
PDFs are plotted in Fig. \ref{fig7} together with fits 
(\ref{nq1}). As is seen from this Figure, in all the cases 
there is a noticeable excess of low mass objects. It can be 
related to the expected disruption of compressed DM walls into 
a system of high density clouds, what decreases the observed 
pancake column density. These results favor a moderate survival 
probability of low mass absorbers. However, such analysis should 
be continued with more precise simulations. 

As was noted in White \& Croft (2003) the formation of LSS elements is
accompanied by their partial disruption and creation of low mass
clouds. This process is more important for formation of galaxies and
their low mass satellites what was also discussed in DD02 and Paper I.
This process can increase the fraction of low mass absorbers in
comparison with theoretical expectations, what, however, can only
partially suppress distortions of the initial power spectrum
considered in this paper.

This discussion confirms that quantitative estimates of the 
shape of the function $\xi_v$ at small scales actually depend 
upon the sample used in the analysis and the model linking the 
observed absorber characteristics with their DM column density 
and the correlation function $\xi_v$. However, differences 
between the measured and the CDM-like correlation functions 
remain quite significant for a wide range of models.

By analyzing the distribution of Lyman-$\alpha$ clouds we can 
test the initial power spectrum of perturbations to unprecedently 
small scale and both outcomes confirmation of the CDM--like shape 
of the power spectrum or detection of distortions are equally 
important. Results of such analysis can be improved with a 
more detailed numerical simulations and a richer sample of 
observed absorbers.  

\section{The initial power spectrum}

For larger scales, $q\geq 0.1$, our approach describes 
reasonably well the properties of richer absorbers and our 
analysis verifies the CDM--like shape of the correlation 
function $\xi_v$ and the initial power spectrum (\ref{f1}). 
The correlation function $\xi_v$ is well fitted by 
the one parameter relation (\ref{xiv}), where the parameter 
$q_0$ is expressed through the spectral moments, $m_0$ and 
$m_{-2}$ (\ref{lv}), and it depends upon the mass of dominant 
DM particles. For $M_{DM}\approx$ 3 -- 5 keV, $q_0\approx 
(1-0.7)\cdot 10^{-2}$ is expected. 

\begin{figure}
\plotone{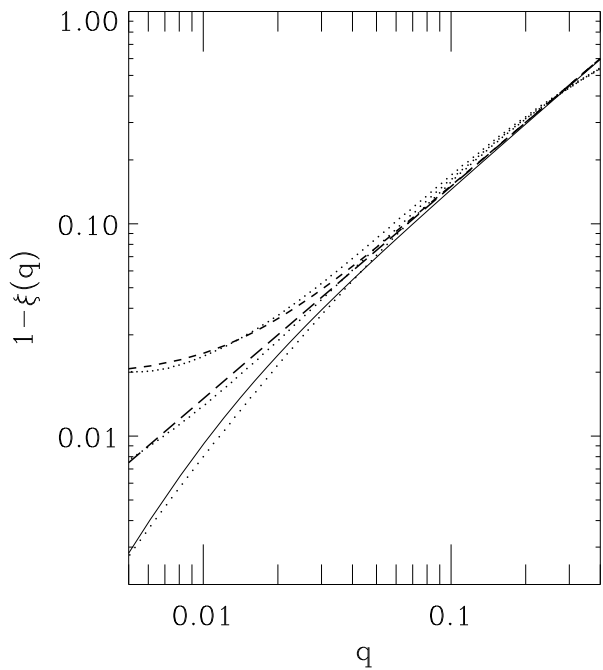}
\vspace{1.5cm}
\caption{Correlation functions for the initial velocity 
field (\ref{xiv}) (solid line) and (\ref{xi400}) 
for the Model 1 and Model 3 (dashed and long dashed lines) together 
with the same functions for spectra (\ref{f11}) and (\ref{disp}) (dotted lines).  
} 
\label{fig8}
\end{figure}

\begin{figure}
\plotone{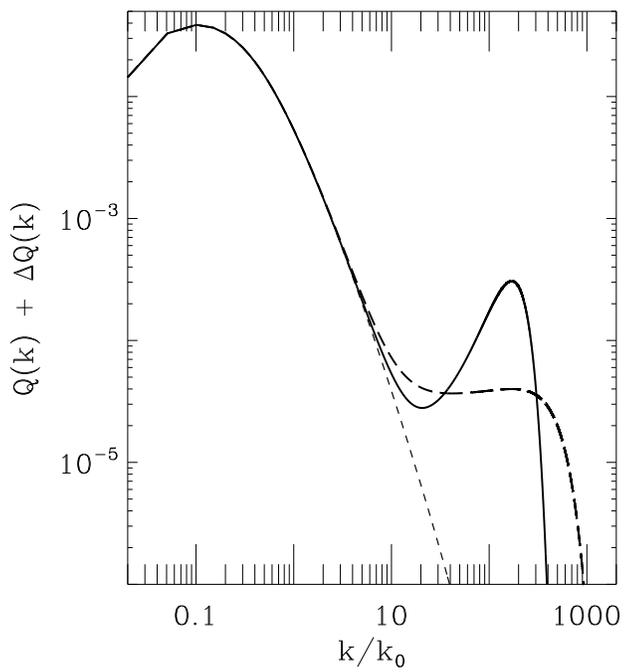}
\vspace{1.5cm}
\caption{Normalized total initial power spectrum 
for two sets of parameters (\ref{disp}) (solid and long dashed lines)
and CDM-like one (\ref{f11}) (dashed line).
} 
\label{fig9}
\end{figure}

However, at small scales the shape of the observed correlation 
function $\xi_v$ differs from the CDM--like one (\ref{xiv}), 
what indicates a more complicated shape of the initial power 
spectrum at small scale. The 
initial power spectrum can be restored from the function $\xi_v$.
Using (\ref{xivv}) we obtain 
\be
P(\eta)\propto \int_0^\infty dq\,q\xi(q)\left[\sin(q\eta)+
2{\cos(q\eta)\over q\eta}\right]
\label{sppc}
\ee
The precision and reliability of our estimates are however limited 
(Sec. 7) and we measure the function $\xi(q)$ only in limited range 
of $q$. So, here we will present only examples of spectra which 
generate the correlation functions $\xi_v$ similar to the measured 
one. As was noted in Sec. 3, the expressions (\ref{qq}) and (\ref{qad}) 
define the DM column density corrected for the Jeans damping and 
therefore as a reference we consider the standard normalized 
spectrum (\ref{f1}) for the mass of dominant fraction of DM 
particles $M_{DM}=3$ keV, $q_0=0.01$: 
\be
Q(\eta)=\eta T^2(\eta)D_W(\eta),\quad \eta=k/k_0,\quad 
D_J(\eta)=1\,.
\label{f11}
\ee
Distortions of the spectrum (\ref{f11}) are described by:
\be
\Delta Q(\eta)={A_p\over\sqrt{2\pi}\sigma_p}
\exp[-0.5(\eta-\eta_p)^2/\sigma_p^2]\,,
\label{disp}
\ee
\[
A_p=5\cdot 10^{-2},\quad \eta_p=170,\quad \sigma_p=65 \,, 
\]
\[
A_p=2.8\cdot 10^{-2},\quad \eta_p=170,\quad \sigma_p=280 \,, 
\]
for Model 1 and Model 3, respectively. The 
correlation functions $\xi_v$ generated by the spectra 
(\ref{f11}) and (\ref{disp}) and these spectra are plotted 
in Fig. \ref{fig8} and Fig. \ref{fig9}\,.

The interpretation of these distortions is not unique because 
of very limited available information. We can estimate the 
function $\xi_v$ only at $q\geq 10^{-3}$ where its dependence 
upon the mass of DM particles is moderate and even for 
$M_{DM}\geq 10^4$ keV it increases of about 1.5 times only. 
This means, for example, that even so essential increase of  
$M_{DM}$ cannot provide noticeable effects and the most probable 
source of the observed distortions are adiabatic or, perhaps, 
isocurvature perturbations predicted by some models of the early 
Universe and inflation.

Recent WMAP measurements confirmed that as expected adiabatic
perturbations dominate on large scale (Peiris et al. 2003)
and that they are Gaussian (Komatsu et al. 2003). However, 
these results do not preclude 
the existence of distortions of the standard CDM--like power 
spectrum at small scales due to both adiabatic and isocurvature 
perturbations (Peiris et al.  2003).

The measured distortions of the power spectrum can be considered 
as an observational evidence in favor of the one field inflation 
with a complicated inflation potential (see, e.g., Ivanov, 
Naselsky \& Novikov 1994) or multiple fields inflation 
(see, e.g., Polarski \& Starobinsky 1995; Turok 1996). 
Both models generate adiabatic or isocurvature deviations from 
the simple CDM--like power spectrum.  More detailed discussion 
of such models can be found, for example, in Peiris et 
al. (2003).

\section{Summary and Discussion.}

In this paper we continue the analysis of a sample of 
$\sim$~4\,500 observed absorbers initiated in Paper 
I and based on the statistical description of Zel'dovich 
pancakes (DD99, DD02). This approach allows one to connect 
the observed characteristics of absorbers with fundamental 
properties of the initial perturbations and demonstrates 
the common origin of absorbers and the Large Scale 
Structure observed in the spatial galaxy distribution 
at small redshifts. 

The physical model of absorbers proposed in Paper I 
and essentially improved in Sec. 3 links three observed 
characteristics of the Lyman-$\alpha$ clouds, namely, the 
redshift, $z$, the Doppler parameter, $b$, and the 
column density of neutral hydrogen, $N_{HI}$, with 
the DM and baryonic column densities, overdensity and 
entropy. It describes redshift dependence of the 
observed characteristics of absorbers discussed in Sec. 
4, discriminates between the random and regular variations 
of absorbers characteristics and allows one to obtain a 
reasonable statistical description of these characteristics. 
It also explains the observed redshift variations of the 
linear number density of absorbers and links them with 
the fundamental parameters of the initial power spectrum. 

Such a complex approach allows one to test its self 
consistency through the comparison of several measured 
statistical characteristics with corresponding theoretical 
expectations. As was shown above, it also allows one to test 
the shape of the correlation function of the initial velocity 
field and of the initial power spectrum on unprecedently small 
scale. 

The progress achieved demonstrates again the key role of 
the representativity and completeness of the observed samples 
for the construction of the physical model of absorbers and 
reveals a close connection between the observational database 
and obtained results. Some of the most interesting results 
were derived from the distribution of weak absorbers but 
unfortunately the available sample of such absorbers has 
only limited statistical significance. Further progress can 
be achieved with a richer sample of observed absorbers and 
by testing our approach on a more representative numerical
simulations.

\subsection{Main results}

Main results of our analysis can be summarized as follows:
\begin{enumerate}
\item We demonstrate that the physical model of absorbers 
        proposed in Paper I and Sec. 3 is self consistent and 
        agrees well with the basic observed characteristics of 
        the Lyman-$\alpha$ clouds. This fact indicates that 
        the majority of absorbers can be considered as 
        long--lived, gravitationally bound and relaxed.  
\item Using the Zel'dovich theory of gravitational instability 
        we link the PDF of DM column density found for the 
        observed sample of Lyman-$\alpha$ clouds with the 
        correlation function of the initial velocity 
        field,  $\xi_v$, what allows one to estimate the 
        initial power spectrum to unprecedently small scale.
\item We found that at larger scales both the measured 
        function $\xi_v$ and the initial power spectrum of
        perturbations do not differ by more than $\sim$ 10--15\% 
	from the predictions of the CDM model. This result 
	confirms the basic predictions of the inflationary
        scenario.
\item At scales $\sim 30-300 h^{-1}$kpc we see noticeable 
        differences between the measured correlation function 
        $\xi_v$ and the expected one for the CDM--like power 
        spectrum. These differences can be considered as an 
        observational evidence in favor of a more complex 
	inflation, which generates an excess of power at small 
	scales. In turn, this excess of power accelerates the 
	formation of first galaxies and causes early reheating 
	of the Universe. 
\item Our quantitative estimates depend upon the limited 
        statistics of weak absorbers and their poorly known 
        survival probability. However, both confirmation 
        of CDM--like shape of the initial power spectrum or 
        detection of its distortions are equally important.  
\end{enumerate}
These results indicate that the proposed approach is quite promising 
and the analysis should be continued with a richer sample of 
observed Ly-$\alpha$ absorbers. Application of this approach to 
simulated matter distribution at $z=0$ and to the sample of $\sim$ 
100\,000 galaxies compiled in the SDSS DR1 demonstrates its high 
efficiency for the quantitative description of strongly nonlinear 
matter condensations. However, its application for the Ly-$\alpha$ 
forest at high redshifts needs to be directly tested with a 
representative numerical simulations what, in turn, will allow 
to improve the estimates of the initial power spectrum.

\subsection{Physical model of absorbers} 

The physical model of absorbers introduced in Paper I 
and improved in Sec. 3 links the measured $z$, $b$ and 
$N_{HI}$ with other physical characteristics of both 
gaseous and DM components forming the observed absorbers. 
In this model we consider the majority of absorbers as 
long--lived gravitationally bound and partly relaxed 
objects formed in the course of both linear and nonlinear 
condensation of the DM component. The existence of galaxies 
and quasars at high redshifts and the observed reheating 
of the Universe at redshifts $z\geq 6$ demonstrate the 
importance of nonlinear processes at redshifts under 
consideration. Numerous simulations (see, e.g., Frenk 2002) 
also confirm the formation of high density DM filaments 
and sheets at these redshifts. 

The Zel'dovich theory of gravitational instability indicates 
the self similar character of the processes of DM structure 
formation and allows formation of numerous 
high density relaxed DM filaments and sheets already at high redshifts. 
Their typical sizes and their mass function depend upon the 
initial power spectrum and their mean characteristics 
progressively change with time. However, the DM structure 
at high redshifts is expected to be qualitatively similar to 
the one observed at small redshifts. The Ly-$\alpha$ clouds
trace this DM structure and, as was shown in Sec. 4, the 
PDFs of their observed characteristics only weakly depend 
on the redshift.  

For low mass pancakes the spatial distribution of the 
baryonic and DM components can be strongly biased due to 
the influence of the gaseous pressure (Jeans damping). 
The physical model of absorbers considered in this paper
allows for this bias. 

Our approach is consistent with the weak redshift variations 
of the mean observed characteristics of absorbers, namely, 
$\langle b\rangle$ and $\langle N_{HI}\rangle$, and even 
the shape of their distribution functions (Sec. 4) does 
not change with time. Moreover, the main statistical 
characteristics of absorbers discussed in Paper I and in 
Secs. 4\,\&\,5 coincide with theoretical expectations. In 
particular, the redshift variations of $\langle\zeta\rangle$ 
do not exceed 10\% (Sec. 4), the mean value of $\zeta$ and 
its measured PDF coincide with the expected ones (Sec. 5.3). 
The same approach allows one to describe the redshift 
variation of the linear number density of absorbers (Sec. 
6). 

Analysis of the main characteristics of absorbers performed 
in Paper I shows that the sample of observed absorbers 
can be composed of DM pancakes with complex evolutionary 
histories. They can be formed due to merging of earlier 
formed pancakes or compression of background matter, and 
both adiabatic, for $\beta\leq 2$, and shock compressions, 
for $\beta\geq 2$ are equally possible. 
The great diversity of these histories is clearly seen 
already from the well known broad distribution of absorbers 
in the $"b-N_{HI}"$ plane (see, e.g., Hu et al. 1995). 

Properties of absorbers are changing even after formation 
of gravitationally bound pancakes. In Paper I we discussed 
five most important factors that determine absorbers 
evolution after formation. They are: the transverse 
expansion and/or compression of pancakes, their merging, 
the radiative heating and cooling of compressed gas, and 
disruption of pancakes into a system of high density 
clouds. To illustrate the influence of these factors 
and correlations between different characteristics of 
absorbers we roughly discriminated in Paper I three 
subpopulations of absorbers with various entropies and 
overdensities. 

Our approach allows one to discriminate between the 
systematic and random variations of properties of absorbers. 
The former ones are naturally related to the progressive 
growth with time of the DM column density of absorbers, 
$q(z)$, what can be described theoretically. The action 
of random factors cannot be satisfactory described by 
any theoretical model. However, in the framework of our 
approach, the joint action of all random factors can be 
described by one random function, which can be identified 
with the entropy of the compressed gas and it is directly 
expressed through the observed parameters (Paper I). 

\subsection{Reconstruction of the initial power spectrum} 

Naturally, the formation and further evolution of DM 
pancakes is determined by the initial power spectrum. 
However, the great diversity of their evolutionary 
histories makes  the reconstruction of this spectrum from 
observations of the Ly-$\alpha$ forest difficult. In this 
paper we solve this problem in the framework of the complex 
statistical investigation of absorbers that allows one to 
check the self consistency of this approach. In spite of 
this, our results are based on some assumptions and, so, 
are model dependent. To check and to improve this approach 
we have  to apply it to the high resolution representative 
simulations of formation of galaxies and absorbers. 

This statistical description connects the PDF of the 
DM column density of pancakes with their survival 
probability, $W_s$, and the correlation function of 
the initial velocity field, $\xi_v$. For Gaussian initial 
perturbations and for a given $W_s$, this connection 
allows one to reconstruct the approximate shape of 
the function $\xi_v$ and to reveal its possible divergence 
from the expected one for the CDM--like initial power 
spectrum corrected for damping due to a finite mass of DM 
particles. In turn, this divergence can be interpreted 
as an indication of existence  of special features in the 
initial power spectrum at small scales. 

In our approach we use the measured PDF of the DM column 
density of pancakes rather than the smoothed density field. 
It therefore avoids the complicated problem of density 
variations along the line of sight and mass conservation. 
It does not depend on the peculiar velocity of absorbers and 
is not restricted by the Nyquist limit and the distances 
between absorbers. So, this indirect approach allows one 
to check the shape of the initial power spectrum in the 
real space up to unprecedently large wave numbers, 
$\eta=k/k_0\sim$ 100. 

For larger scales, $\geq 1-2h^{-1}$ Mpc,\, $\eta\leq 3$, our 
analysis verifies the CDM--like shape of the initial power 
spectrum and the correlation function $\xi_v$. However at 
small scales noticeable differences between the CDM--like 
and measured spectra are detected. The reliability of these 
differences is problematic (see Sec. 7) but both confirmation 
and/or elimination of these distortions are equally important. 

\subsection{Numerical simulations and the model of "fluctuating 
Gunn--Peterson approximation"} 

The leading role of numerical simulations in investigation of 
the LSS is well known. They clarify many problems of structure 
formation and evolution at all redshifts and, in particular, 
link the observed characteristics of absorbers with those that are not
directly observable. 
This link allows one to test the physical model of absorbers 
and, further on, to use it to recover the density fields and 
the initial power spectrum from the observed Ly--$\alpha$ forest. 
Here we shortly compare our results with numerical simulations.

\subsubsection{The model of "fluctuating Gunn--Peterson approximation"} 

One of the popular model of the observed Ly--$\alpha$ forest is 
the "fluctuating Gunn--Peterson approximation" (FGPA) widely discussed 
during the last five years (see, e.g., Croft et al. 1998, 2002; 
Nusser\,\&\, Haehnelt 2000; McDonald et al. 2000; Phillips et al. 
2001; Zaldarriaga, Hui \& Tegmark 2001; Shaya et al. 2002). It is 
focused on the connection between the observed redshift variations 
of the flux emitted from quasars and the initial power spectrum.  

In this approach absorbers are considered as formed by a mildly 
nonlinear unshocked compression of the background matter. It 
links absorption lines with large diffuse gaseous clouds with 
moderate overdensities which are expanded by the residual Hubble 
flow. For such clouds their velocity profile depends mainly on 
their size, while the thermal broadening is considered as minor 
(Croft et al. 2002). However, other authors consider this  
broadening as quite essential (see, e.g., Meiksin, Bryan \& 
Machacek 2001; White \& Croft 2003). 

The main result of the FGPA is the determination of the power 
spectrum. This method does not use the observed hydrogen column 
density and the Doppler parameter. It begins with the analysis 
of the measured one dimensional spectrum of the flux and 
converts it into the power spectrum of the matter. The nonlinear 
effects are partly taken into account when the measured flux 
spectrum is converted to the matter spectrum with the special 
function $b(k)$ (Croft et al. 2002). This simple method, in 
spite of its deficiencies, surprisingly well restores the 
CDM--like power spectrum down to scales $\sim 1h^{-1}$Mpc. 

The application of the FGPA is restricted by the Nyquist limits. 
For the typical separation of observed absorbers (\ref{tmns})
\be
\langle D_{sep}\rangle\approx 1 \left({0.2\over\Omega_m h}
\right)h^{-1}{\rm Mpc}\,,
\label{nyq}
\ee
the analysis allows one to reach a typical wave number 
\[
k_N\approx {\pi\over 2D_{sep}}\approx 1.6\left({\Omega_m 
h\over 0.2}\right) h{\rm Mpc}^{-1},\quad \eta_N={k_N
\over k_0}\approx 8\,.
\] 
At such scales, the similarity of the observed spectra with 
the CDM--like one is confirmed by our results (\ref{xi11}).

Comparison of simulated and measured flux characteristics 
in McDonald et al. (2000) shows an excellent agreement. However, 
later analysis revealed systematic inconsistencies between the 
observed and simulated intermittent features of the flux 
that indicates the underestimation of nonlinear evolution 
in simulations (Pando et al. 2002). White \& Croft (2003) 
note also the importance of the non--linearity of matter 
clustering for the reconstruction of the power spectrum. 

The FGPA has been criticized, for example, by  Zaldarriaga, 
Scoccimorro \& Hui (2002) who notice the limited usefulness 
of available simulations, limitations of the inversion technique 
and the importance of nonlinear effects. The same problems are 
also discussed in Gnedin \& Hamilton (2002), Tegmark \& 
Zaldarriaga (2002) and Seljak, McDonald \& Makarov (2003). 

\subsubsection{Properties of simulated forest} 

Usually the comparison of observed and simulated characteristics 
of the Ly--$\alpha$ forest is restricted by the flux emitted from 
quasars but always they are found to be quite consistent. However, 
the flux integrates the HI distribution over absorber and therefore 
it does not characterize it uniquely.

More detailed comparison of the Doppler parameter, $b$, and the 
column density of neutral hydrogen, $N_{HI}$, performed in Meiksin, 
Bryan \& Machacek (2001) reveals that no simulations provide a 
statistically acceptable match to the observed $b$ and $N_{HI}$ 
distributions. In particular, they found that the median measured 
Doppler parameters exceed the simulated one by as much as 30 -- 60\% 
especially for optically thin lines and also the simulated and 
measured $N_{HI}$ distributions tend not to agree.
It can be expected that results presented in Sec. 4.2 enhance 
divergences between the simulated  and observed {\it quantitative} 
characteristics of the forest. 

As was noted in Paper I, for the sample dominated by recently 
formed 'young' absorbers a strong redshift evolution of observed 
characteristics and, in particular, of the Doppler parameter, is 
expected. Perhaps, this effect enhances distortions between measured 
and simulated Doppler parameters.

The available numerical simulations are restricted by technical 
limitations but, in spite of this, they could help to solve many 
unclear problems of the early structure formation. First of all, 
it is link between the gaseous and DM components and estimate of 
the degree of nonlinearity. A detailed description of the DM 
distribution is now possible with at least two effective methods 
-- the Minimal Spanning Tree (Demia\'nski et al. 2000; Doroshkevich 
et al. 2001) and the Minkowski Functional (see, e.g., Schmalzing 
et al. 1999) techniques. Characteristics of DM distribution are 
extremely important for the reconstruction of the processes of 
formation of the Ly-$\alpha$ forest. These characteristics can 
be directly compared with expectations of the Zel'dovich theory 
discussed in Sec. 2\,.

Now it is not clear why the FGPA links the Ly-$\alpha$ forest 
with large diffuse gaseous clouds with moderate overdensities 
and why the high overdensity DM pancakes -- relaxed or unrelaxed 
-- are not seen in simulations. It is natural to expect that the 
formation of galaxies is accompanied by formation of numerous 
high overdensity pancakes. Simple estimates show that, for 
parameters discussed in Sec. 4.2, the one dimensional collapse 
leading to formation of high density pancakes is quite probable 
at all redshifts under consideration. For $\langle b\rangle/
b_{bg}\geq\beta_{thr}$ shock compression dominates. 

It is also not clear what happens with the formed DM clouds 
in the course of their further evolution and why merging of 
pancakes is suppressed. It is one of the important processes 
responsible for evolution of the structure which is clearly 
seen in numerous high resolution simulations performed in large 
boxes. This process is accompanied by shock compression of the 
gas and an essential growth of its entropy. In turn, variations 
of the entropy explain the strong scatter of absorbers in the 
"$b-N_{HI}$" plane (Hu et al. 1995). This list of questions can 
be substantially extended. 

\begin{figure}
\plotone{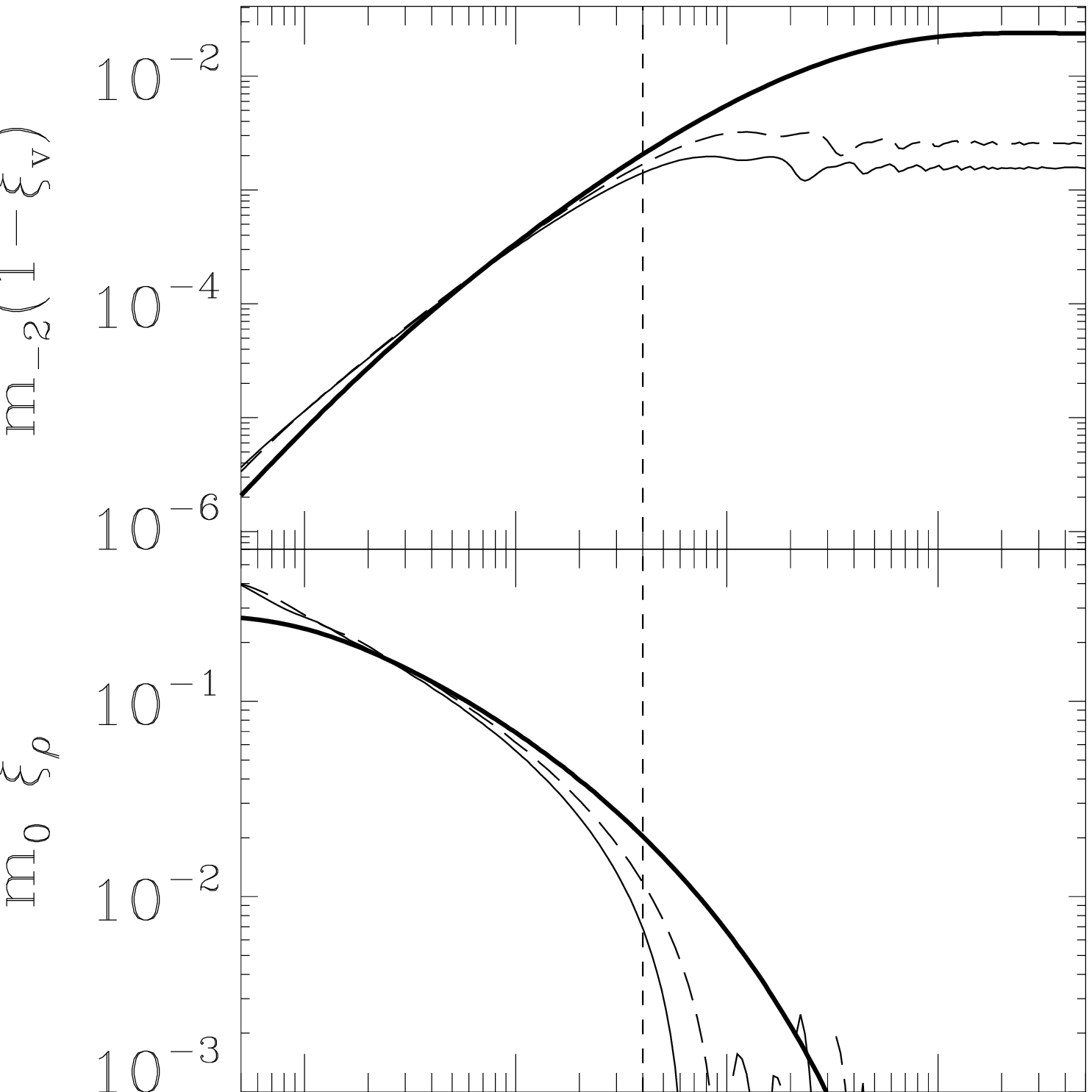}
\vspace{1.5cm}
\caption{Normalized correlation functions of initial velocity 
and density fields, $\xi_v\,\&\,\xi_\rho$, vs. the dimensionless 
point separation, $q l_v k_0$, for the CDM--like power spectrum 
(thick solid line) and for truncated spectra used in simulations 
with $256^3$ cells and the box sizes $k_0 L_{box}=2.1$ (thin solid 
line) and 3.14 (long dashed line). 
} 
\label{fig10}
\end{figure}

To achieve the required resolution, the simulations of the 
forest are usually performed in relatively small boxes with 
sizes $L_{box}\sim 5-20 h^{-1}$Mpc what eliminates the large 
scale initial perturbations. 
To test how accurately such numerical simulations reproduce
observations in Fig. \ref{fig10} we compare the velocity correlation 
function $\xi_{v}$
for CDM power spectrum with velocity correlation functions for two
truncated spectra used in simulations. 
As is seen from this Figure, both correlation 
functions derived from truncated spectra at scales $l_v q k_0\geq 0.4$
are suppressed. This in turn slows down
the merging of clouds, decelerates the 
formation of richer absorbers and distorts their other characteristics. 
As was found in Secs. 4.2 and 5.3, for the cosmological model 
(\ref{basic}) the mean size of compressed absorbers is 
\be
l_v k_0\langle q(z) \rangle\approx 
{1.1 l_vk_0\langle\zeta\rangle\over (1+z)^2}\approx 
0.4\left({4\over 1+z}\right)^2\,,
\label{qk0}
\ee
what coincides with the critical value found above. 

This 
estimates indicate that in simulations performed with $k_0L_{box}
\leq 5 - 6$ the formation of richer absorbers with $q\geq\langle 
q(z)\rangle$ is strongly suppressed and the weaker adiabatically 
compressed absorbers dominate. On the other hand, estimates 
discussed in Sec. 4.2 verify that at redshifts under consideration at least 
half of matter could be already accumulated by absorbers and, so, 
their evolution could be driven by the processes of merging and shock 
heating. These inferences should be tested on  simulated DM characteristics 
of pancakes and compared with expected ones for the corresponding 
power spectrum, for example, with methods described in Sec. 2\,. 

\subsubsection{Measurements of the power spectrum} 

The  FGPA picture seems to be quite different from our 
representation of absorbers as a network of discrete 
gravitationally bound pancakes with a variety of overdensities. 
Our model identifies the Doppler parameter with the depth of 
the DM potential well and the temperature of compressed gas
rather than with the residual expansion of matter. It assumes 
domination of long lived absorbers what agrees well with results 
obtained in Sec. 4.2. Direct estimates obtained with the pancake 
sizes and Doppler parameters specified in this section show that 
the formation of such high density absorbers is quite probable 
at all redshifts under consideration.

Both models describe quite well some of the observed properties of the
Lyman-$\alpha$ forest but they differ in emphasizing domination of
discrete or diffuse absorbers. In many respects they are complementary
to each other because diffuse and discrete absorbers coexist and
fractions of these types of absorbers change with redshift. Evidently
the diffuse absorbers dominate just after reheating while discrete
absorbers dominate at lower redshifts.  These differences demonstrate
the complex character of the Lyman-$\alpha$ absorption clouds and
indicate that both models should be improved after detailed analysis
of truly representative simulations of the forest.

The proposed here method of measuring the power spectrum is based 
on the strong connection between the power spectrum and the mass 
function of absorbers and seems to be quite promising. It is not 
restricted by the Nyquist limit and therefore it can be 
extended to scales about ten -- thirty times smaller than the 
method based on the FGPA. It is not rigidly linked to the accepted 
physical model of absorbers which can be corrected and improved 
in the course of further investigations. It can be also easily 
generalized for more refined relations between the mass function 
and correlation function of velocity field. These corrections can 
be introduced  after detailed investigations of improved and truly 
representative simulations which reproduce {\it all} observed 
characteristics of the Ly-$\alpha$ forest and link them with DM 
characteristics of pancakes. 

\subsection{Reheating of the Universe and the initial 
power spectrum}

Recent observations of high redshift quasars with $z\geq$ 5 
(Djorgovski et al. 2001; Becker et al. 2001; Pentericci et 
al. 2001; Fan et al. 2001, 2003) provide a clear evidence 
in favor of the reionization of the Universe at redshifts 
$z\sim$~6, when the volume averaged fraction of neutral 
hydrogen is found to be $f_H\geq 10^{-3}$ and the 
photoionization rate $\Gamma_\gamma\sim (0.2 - 0.8)\cdot 
10^{-13}{\rm s}^{-1}$ . These data coincide with those expected 
at the end of the reionization epoch which probably took 
place at $z\sim$~6. On the other hand, recent WMAP data 
shift the redshift of reionization up to $z\sim 20\pm 10$ 
(Kogut et al. 2003), what implies a complex ionization 
history. 

The reionization at $z\sim 6 - 10$ is consistent with 
the CDM--like power spectrum with the mass of dominant 
DM particles $M_{DM}\sim 1 -5$ keV (see discussion in Paper 
I). Earlier reionization is compatible with a larger mass 
of the dominant fraction of DM particles, $M_{DM}\geq 1$ 
MeV or with distortions of the CDM--like initial power 
spectrum at small scales, what in turn implies a complex 
inflation and a more efficient formation of galaxies at high 
redshifts (see discussion in  Spergel et al. 2003). 

Models with complex inflation and an excess of power 
at small scales were recently discussed in connection with 
distortions of the CMB polarization  (Naselsky \& 
Novikov 2002; Doroshkevich, Naselsky~I., Naselsky~P., 
Novikov 2003)
and galaxy formation (Sommer--Larsen 
et al. 2003). This means that measurements 
of the initial power spectrum at small scale are important 
for many aspects of modern cosmology. 

The future Planck mission will provide high accuracy 
measurements of the CMB polarization and clarify the 
ionization history of the Universe. However, these 
results will give again only indirect information about 
the possible distortions of the power spectrum. This 
means that the measurement of such distortions with 
the Ly-$\alpha$ forest remains the most efficient 
way for investigation of the initial power spectrum
at small scale. 

The preliminary estimates obtained above seem to be 
quite promising but must be tested with richer samples 
of observed absorbers. Such investigations will allow 
one to obtain more detailed and reliable information and 
better understanding of the physical processes at high
redshifts and the shape of the initial power spectrum.

\section*{Acknowledgments}
AGD is grateful to Dr. S. Cristiani and Dr. T.S.Kim for 
the permission  to use the unpublished observational data.  
This paper was supported in part by Denmark's
Grundforskningsfond through its support for an establishment 
of Theoretical Astrophysics Center and by a grant of the 
Polish State Committee for Scientific Research. 
MD wants to thank TAC for great hospitality and support. 
AGD also wishes to acknowledge support
from the Center of Cosmo-Particle Physics, Moscow.
Furthermore, we wish to thank the anonymous
referee for many useful comments.

{}
\clearpage

\clearpage

\clearpage

\clearpage

\clearpage

\clearpage

\clearpage

\clearpage
    
\clearpage
 
\clearpage
 
\clearpage

\begin{table}
\caption{QSO spectra used in our analysis}
\label{tbl1}
\begin{tabular}{ccccr}
    &$z_{em}$&$z_{min}$&$z_{max}$&No \\
               &     &   &   &of HI lines\\
$0000-260^{1}~$ & 4.11&3.4&4.1& 431\\       
$0055-259^{2}~$ & 3.66&3.0&3.6& 534\\       
$0014+813^{3}~$ & 3.41&2.7&3.2& 262\\       
$0956+122^{3}~$ & 3.30&2.6&3.1& 256\\       
$0302-003^{3, 2}$&3.29&2.6&3.1& 356\\       
$0636+680^{3}~$ & 3.17&2.5&3.0& 313\\       
$1759+754^{4}~$ & 3.05&2.4&3.0& 307\\       
$1946+766^{5}~$ & 3.02&2.4&3.0& 461\\       
$1347-246^{2}~$ & 2.63&2.1&2.6& 361\\       
$1122-441^{2}~$ & 2.42&1.9&2.4& 353\\       
$2217-282^{2}~$ & 2.41&1.9&2.3& 262\\       
$2233-606^{6}~$ & 2.24&1.5&2.2& 293\\       
$1101-264^{2}~$ & 2.15&1.6&2.1& 277\\       
$0515-441^{2}~$ & 1.72&1.5&1.7&~~76\\       
\vspace{0.15cm}
\end{tabular}                               

1. Lu et al. (1996),
2. unpublished, courtesy of Dr. Kim   
3. Hu et al., (1995),
4. Djorgovski et al. (2001)
5. Kirkman \& Tytler (1997),
6. Cristiani \& D'Odorico (2000).
\end{table}

\end{document}